# An in-depth Analysis of Occasional and Recurring Collaborations in Online Music Co-creation

Fabio Calefato, Giuseppe Iaffaldano, Leonardo Trisolini, Filippo Lanubile
University of Bari, Italy

The success of online creative communities depends on the will of participants to create and derive content in a collaborative environment. Despite their growing popularity, the factors that lead to remixing existing content in online creative communities are not entirely understood. In this paper, we focus on *overdubbing*, that is, a dyadic collaboration where one author mixes one new track with an audio recording previously uploaded by another. We study musicians who collaborate regularly, that is, frequently overdub each other's songs. Building on frequent pattern mining techniques, we develop an approach to seek instances of such recurring collaborations in the Songtree community. We identify 43 instances involving two or three members with a similar reputation in the community. Our findings highlight common and different remix factors in occasional and recurring collaborations. Specifically, fresh and less mature songs are generally overdubbed more; instead, exchanging messages and invitations to collaborate are significant factors only for songs generated through recurring collaborations whereas author reputation (ranking) and applying metadata tags to songs have a positive effect only in occasional collaborations.

## 1 INTRODUCTION

Online communities allow people with a shared purpose or interest to interact remotely and share content in the same online environment [43]. Participants in online communities have succeeded in developing huge knowledge bases (e.g., Wikipedia), large software projects (e.g., Linux Kernel, Mozilla Firefox, R), and creative work including videos, digital animations, and music (e.g., Deviant Art, Newgrounds).

There is a large body of prior work that has investigated collaboration in different types of online communities where, due to the differences in the artifact of interest and community characteristics, the findings often vary. In this study, we focus on investigating collaboration from the perspective of creators, i.e., members who actively share artifacts (e.g., writing songs) in peer-production communities, as opposed to non-creators, i.e., end-users who participate in the community exclusively by consuming them (e.g., playing songs). Albeit both creators and end-users are necessary for the survival of online peer-production communities, the presence of the latter is consequential to the availability of shared artifacts. Accordingly, we look at collaboration in online creative communities in terms of artifact *remix* (or *reuse*),[1] whereby community members generate

https://doi.org/10.1145/3493800

---
[1] Hereinafter, the terms artifact *remix* and *reuse* are used interchangeably.

derivative content through the reworking and recombination of existing contributions shared by others [11,28,40].

Despite the considerable amount of existing research on creative communities, the factors that lead to the reuse of specific artifacts over others are not entirely understood [46]. For instance, Luther et al. have questioned whether the factors related to the propensity for reusing are artifact- or domain-dependent [34]. Previous research in open source software (OSS) communities has established that the contribution of source code changes by OSS developers depends on both social and technical factors [22,23,51]. Also in the case of creative arts communities, Luther et al. [33,34] found that the social reputation of participants is key to completing collaborative animation efforts. Burke and Settles [6] found that users, especially newcomers, who engage in social features and one-to-one collaborations achieve their songwriting goals better than those who are non-social. Other studies, instead, focused on factors that lead members of arts communities to select specific creative artifacts shared by others for reworking and recombining them into something new [11,28,50].

To further our understanding of the factors influencing the reuse of existing content, we build on prior work on collaboration in online creative communities (including OSS development) to design a comprehensive study on Songtree, an online community for music co-creation. We focus on the creative action of *overdubbing*,[2] a form of dyadic collaboration whereby a new track is mixed with an existing audio recording (e.g., singing over an instrumental song) previously uploaded by others.

We analyze a dataset of 263K songs and 57K authors extracted from the Songtree database. We perform a sophisticated regression study to analyze the relationship between the song- and author-related measures (e.g., likes, followers) and the probability of songs being overdubbed as well as the count of overdubs received. Overall, we find that both recent and less mature songs as well as those that receive likes, bookmarks, reposts, and technical specification tags are more likely to be reused at all and receive a higher number of remixes. Furthermore, we find evidence that songs by more popular authors (i.e., with a high reputation and many followers) and whose songs are often reused have a higher probability of being remixed.

Furthermore, when analyzing overdubs, we observe that some of these pairwise collaborations are recurring, that is, we find several instances of collaborations where author A overdubs songs from author B, who in turn extends some of the recordings uploaded by A. Previous work on online creative communities has devoted surprisingly little attention to studying recurring collaborations, considering the co-creation of content almost exclusively as an occasional endeavor. Recurring 'collabs' in online communities for music co-creation are mentioned in the work on FAWM by Dow and Settles [15], where they report the case of three members who formed a virtual band to compose 42 songs about the U.S. presidents and adopted a *"parallel, distributed-labor model of collaboration reminiscent of open-source software […] projects."* Previous research on OSS development has proven the existence of co-development groups, i.e., latent socio-technical structures [4] formed by two or more developers (also referred to as *implicit teams* [38] and *putative groups* [20]) who tend to communicate often and work together on the same artifacts. There is also evidence that the members of these co-development groups are very productive and more likely to remain active within communities for longer [56]. Consistently, theories on group attachment suggest that groups who 'interact' (i.e., work and talk) rather than just 'coact' (i.e., work together without interaction) remain active longer and reach higher performances [6]. Understanding and fostering frequent collaborations would help creative community managers ensure that members remain productive for a longer period to sustain actively the community. Accordingly, in this paper, we analyze recurring collaborations to understand (i) how frequently they occur, (ii) the characteristics of their members, and (iii) the remix factors as compared to those identified in occasional collaborations. We also conduct an expert consultation session with the Songtree founder, to garner additional insight and triangulate the findings.

---

[2] *Overdubbing* is also intended as a form of remix/reuse applied to music artifact.

To mine recurring collaborations, we devise a new algorithm – adapted from an existing approach for frequent pattern mining – and uncover the sets of authors who frequently connect through overdubbing. We identify 43 instances of recurring collaborations involving two or three members. We find that the instruments and genres played in occasional and recurring overdubs only partially overlap. We also confirm previous findings of collaborations being associated with a small delta in the community reputation between the parties involved. Finally, regarding the remix factors (also referred to as antecedents of reuse), we find that overall, they are similar between occasional and recurring collaborations, but the number of messages exchanged between the two parties and sending invitations to overdub are both significant antecedents of reuse only for the latter.

**Novelty**. This study largely extends our earlier work [7,8] where we began studying remix factors in online music communities. Most notably, the analysis of recurring collaborations is original. Also, while we carry part of the hypotheses and their operationalization from our previous work, here we add three more hypotheses about the effect of song metadata and member interaction. In addition, we develop a couple of more sophisticated count data models, which combine a linear model and a logistic model, thus enabling a more refined analysis of the antecedents of reuse.

**Contributions.** This paper makes the following main contributions. First, from a research perspective, we study reuse in the Songtree music community, which has been considerably less investigated than other online creative communities, such as Scratch and Newgrounds, thus adding further evidence to the existing body of knowledge. Second, we investigate the extent to which recurring collaborations happen within Songtree, by adapting and applying an algorithm based on frequent pattern mining, and derive a taxonomy of recurring collaborations. Third, we propose the use of the signaling theory as a framework for interpreting the results.

Furthermore, from a practical perspective, some of the antecedents of reuse identified through our regression analyses are actionable and, therefore, can be acted upon by Songtree users who want to increase their community status when uploading their artwork: remix fresh and less complete content as well as use tags to provide technical specifications (e.g., instruments played, tempo) of uploads. We also make some practical recommendations to the designers of online platforms for music co-creation; our analysis of recurring collaborations reveals several shortcomings and lack of collaborative features that are instead considered commodities in other online collaborative platforms such as those for hosting OSS projects (e.g., GitHub and GitLab).

**Structure of the paper**. The remainder of this paper is organized as follows. In Section 2, we describe Songtree and its key concepts, which we use afterward to design our empirical analysis. In Section 3, we review prior work on collaboration in online creative communities, then we define the concepts related to recurring collaborations. In Section 4, we describe our research framework organized in two stages, the first one to test the hypotheses related to the antecedents of song reuse in occasional collaborations and the second to answer the research questions related to the mining of recurring collaborations. Section 5 describes the Songtree dataset and the measures extracted. Section 6 reports the results from the two stages of our empirical investigation. Findings are discussed in Section 7, along with the limitations of our study. Finally, we conclude in Section 8.

## 2    SONGTREE

Songtree[3] is an online creative music community, grown upon a collaborative software platform, where artists participate in the creation of songs. As of November 2019, the community counted about 295K registered users, of which ~57K are authors who uploaded over 263K songs. Songtree allows users to extend (namely, *overdub*) any publicly shared song without permission by mixing (i.e., adding) one additional track. This process is non-destructive, as the original song does not change; what happens instead is that a new version of the song is created and linked to the original. Songtree leverages the metaphor of a growing tree to represent and keep track of the collaborative creation of a song (see Figure 1). In the example, user John uploads a new song (step #1), which becomes the root of the song tree (the topmost node). Then, user Paul records a guitar track over John's song (step #2), thus creating a new node (overdub) branching out from the root node. Because songs are public, while John and Ringo further extend Paul's version (steps #3 and #4), members with a different taste in music can create other versions of John's original root song (step #5), taking different directions. Thus, over time, the tree of a song gradually grows as new overdubs are posted, each representing an extension of its parent song.

---

[3] http://songtr.ee

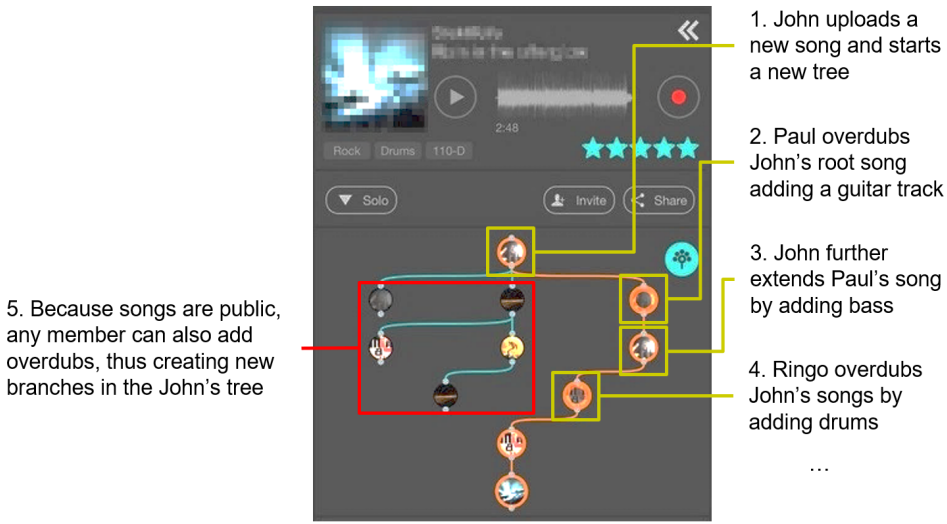

Figure 2. A fictional example of collaboration in a song tree.

Songtree's collaboration workflow is in close analogy to the modern workflow typically used in OSS projects. The so-called *fork-and-pull model* popularized by GitHub[4] is common among OSS projects as it reduces the amount of friction for new contributors and allows developers to work independently without upfront coordination. In this model (see Figure 2a), anyone can *fork* an existing software repository (i.e., copy a project to their personal space) and push changes to their fork without needing access to the source repository. Then, changes can be *pulled* into the source repository by opening a pull request to propose the updates from one fork to be integrated into the original source repository. In Songtree (see Figure 2b), overdubbing is the equivalent of forking in GitHub, as any song from one author can be extended without permission by others and saved to their personal space; however, Songtree lacks a mechanism equivalent to pull requests to ask permission from the original author for adding the overdub's new track to the original song.

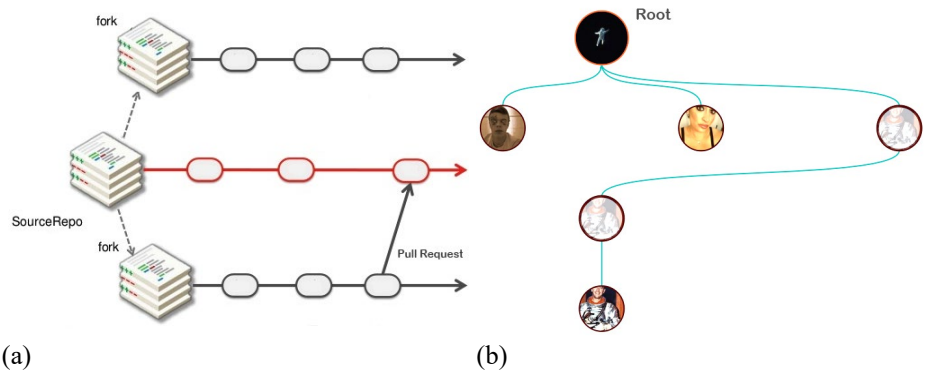

(a)                                          (b)

Figure 1. A comparison between collaborative workflows in GitHub (a) and Songtree (b).

---

[4] https://guides.github.com/introduction/flow

Songtree offers several social-networking features such as the ability to follow other musicians, as well as to like, bookmark, and repost songs. Songs receiving a high number of reactions from the community may also be showcased in Songtree under the *Popular* section of the website; In addition, to favorite the serendipitous discovery of new artists and songs, Songtree includes the *Latest*, *Featured*, and *Top artists* sections, listing songs and members according to the release time, an internal recommender system, and community ranking. Finally, Songtree users can earn badges through their activity within the community. There are three badge categories, namely *new songs*, *overdubs*, and *overdubs received*, earned respectively by uploading new own songs, overdubbing other songs, and receiving overdubs by others. In Songtree, badges act as a proxy measure of user reputation within the community, measured by the quantity and quality of the content created therein. Finally, user profile pages list authors' personal information, biography, pictures, followers, external links, and statistics of their activity.

## 3  COLLABORATION IN ONLINE CREATIVE COMMUNITIES

In line with the work of Luther et al. [34], in our study, we focus on studying collaboration in the form of overdubbing, that is creating a remix (extension/reuse) of an existing song. Regarding the remix factors, they are intended along the lines of Cheliotis et al. [11] as the antecedents influencing the propensity to overdub an existing song. In the following, we first review the findings from prior work to identify remix factors related to collaboration in online creative communities (Sect. 3.1). Then, we review prior work focusing specifically on recurring collaborations (Sect. 3.2).

*3.1 Remix Factors*

Because online creative communities vary over a wide spectrum depending on the type of members, the shared purpose that drove them together, and the type of artifact generated therein [36], here we review how remix factors reported in prior work tend to vary. Our goal here is to identify and adapt the concepts used in previous work to assess whether existing findings generalize to online communities for music co-creation such as Songtree.

In creative communities, members produce artifacts that require a great degree of creativity or artistic skills [46]. Collaboration in creative arts communities typically happens through reuse, i.e., the generation of derivative content through the reworking and recombination of existing contributions such as music, 3D arts, and animations [11,28]. In music communities, in particular, reuse is mostly referred to as remix, where it indicates "*a reinterpretation of a pre-existing song*" [40]. Nevertheless, the term remix has now become common also in creative contexts other than music and is therefore used for referring to, for example, reusing video animations [28] and 3D-printable content [50]. Cheliotis et al. [11] investigated the likelihood of songs being remixed in the ccMixter music community. They found that the degree of derivativity (i.e., remixes 'closer' to their parent songs), fecundity (i.e., being an author with a history of remixes received), and social embeddedness (i.e., having a high level of commitment and contribution to the community) are all positive antecedents of remixes. Hill and Monroy-Hernández [28] performed a study on Scratch, an online community where amateur creators combine images, music, and sound to obtain Adobe Flash-like video animations. They found that the likelihood of engendering derivative works is related to work complexity and author prominence. Also, in direct contrast with the finding by Cheliotis et al. [11] on the degree of derivativity (i.e., the 'newer' the content, the higher the likelihood of remix), Hill and Monroy-Hernández found support for their hypothesis about work cumulativeness, observing that remixes are more likely to be reused than *de novo* content. Stanko [50] investigated why some 3D-printable objects in the Thingiverse community are more generative than others. He found that the remixing likelihood is positively related to the interaction with other community members. Luther et al. [33,34] investigated the role of leadership and other factors influencing the success of collaboration in Newgrounds, a collaborative animation community. In [34] they found that the collaborations more likely to be completed are those initiated by experienced 'leaders' well-known to the community (as the number of views and likes received by animations helps build up their reputation), who are also inclined to communicate frequently. In another study on Newgrounds [35], they also found evidence that specifying technical constraints, such as the frame rate and background color of the animation, in collaboration descriptions is associated with a higher chance of their success. Settles and Dow [46] analyzed FAWM (February Album Writing Month), an online community for songwriters who collaborate every year to the creation of an entire album of songs in one month. They found evidence that prior interactions (i.e., the exchange of direct messages) and having a small delta in a community's social reputation are key factors in pairing; also, the perception of balanced efforts from both parties is the factor that contributes the most to the completion of such collaborations.

In conclusion, prior work has highlighted the existence of both technical and social factors as antecedents of reuse—in our case, the factors related respectively to songs and authors in online music co-creation. Regarding the technical aspects, our review has identified that artifact metadata (e.g., music tempo, song length in our case) and time (e.g., when an overdub was recorded) are potential remix factors to consider. Concerning the social aspects, our review of existing findings suggests that user reputation (e.g., followers) and artifact-related feedback (e.g., number of likes, plays) are also potential antecedents of reuse to include in our analysis.

*3.2 Recurring Collaborations*

In this section, we review prior works on recurring collaboration in online creative communities, with a particular focus on music co-creation and software development.

There is a surprisingly limited amount of previous research on recurring collaborations in online artistic communities. Settles and Dow [46] and Dow and Settles [15] studied the factors influencing the formation of collaborations (or *collabs*) in the FAWM music community. They observed that collabs form out of shared interests but different skills and involve members with small differences in their community ranking. Also, they noted that communication exchanges are predictive of collab formations. Interestingly, albeit one-to-one, pairwise collaborations were predominant in FAWM, they mention the case of a collab involving three community members who worked together following a "*parallel, distributed-labor model of collaboration […] reminiscent of open-source software*" (see [15], p. 22). The three members ended up starting a band that released a triple album and toured music festivals. Silva et al. [48] studied how professional musicians collaborate and how such connections impact their music success (i.e., Billboard ranking). Their main findings are that successful artists have a high degree of connections and diversification as collaborations help them to bridge gaps between styles and genres, and cross over to new fan bases. In follow-up work, Silva and Moro [47] were able to establish the presence of a causal relationship between collaboration and success.

Concerning software development, extensive research has been conducted on the formation of virtual teams [21], particularly in domains such as e-learning [53] and global software engineering [19]. However, our focus here is not on investigating established teams but rather on groups of individuals who spontaneously get together and collaborate recurrently, acting *de facto* as a team. Xuan and Filkov [57] and Gharehyazie and Filkov [20] investigated synchronous group co-development within the Apache software ecosystem. They identified the presence of *putative collaborative groups* (CoG) of developers who act like teams since they tend to work together in code proximity (i.e., modify the same source files) and time proximity (i.e., around the same time). They observed that the activity of these putative groups results in commit bursts (i.e., more lines of code added), which are in turn associated with communication bursts (i.e., more email exchanged), required to synchronize the collaboration. Finally, they found CoGs of size two to be much more prevalent than larger groups of size three or more.

Overall, from the review above, it is clear the presence of a limited amount of research on recurring collaboration in the field of online creative communities, especially artistic, which we aim to further with this work.

## 4   RESEARCH FRAMEWORK

Since we aim to investigate the remix factors associated with overdubbing in Songtree, we devise a research framework divided into two stages (see Figure 3). In the first research stage, which we refer to as occasional collaborations, we replicate and extend the results of our prior studies on remix factors associated in general with overdubbing songs. In the second research stage, we investigate the presence of recurring collaborations and compare the remix factors of songs written by frequent collaborators against the antecedents of reuse identified in the first stage. Further details about each workflow are given next and in Section 5.

### 4.1 Stage 1: Occasional Collaborations

In the first stage of our research framework, we elaborate and test a set of eight hypotheses built upon prior work and the observations obtained from a couple of sessions conducted with the Songtree development team. Five of these hypotheses (numbered *H1-5*) are carried over from our prior studies on remix antecedents [7,8], and the related findings are summarized below in Sect. 4.1.1. The remaining three hypotheses (numbered $\overline{H}$*6-8*) are novel and introduced in Sect. 4.1.2.

**4.1.1 Summary of Findings from Prior Work on Songtree**

In our prior work [7,8], we developed a set of five hypotheses to analyze the relationship between song- and author-related measures and the likelihood of remixing songs in Songtree.

Regarding the first hypothesis *H1* (*The number of reactions generated by songs is positively associated with receiving overdubs*) we found that the number of likes, bookmarks, and reposts (i.e., positive feedback) received by a song is positively associated with its likelihood of being overdubbed. For *H2* (*Time is negatively associated with receiving overdubs*), we found that the time since the upload of a song is negatively associated with its likelihood of being overdubbed at least once – that is, songs that do not receive the first overdub soon after being uploaded will likely never be remixed at all. As for *H3* (*The degree of derivativity of songs is negatively associated with receiving overdubs*), we observed that the distance of a song from the root of its tree is negatively associated with the likelihood of being overdubbed at all – in other words, the more derivative a song is, the fewer its chances to be remixed. Regarding *H4* (*The ranking of authors in the community is positively associated with receiving overdubs*), community reputation and the gamification badges gained by being overdubbed are associated with higher odds of receiving further. Finally, we found confirmation for hypothesis *H5* (*Customizing the author profile avatar is positively associated with receiving overdubs*).

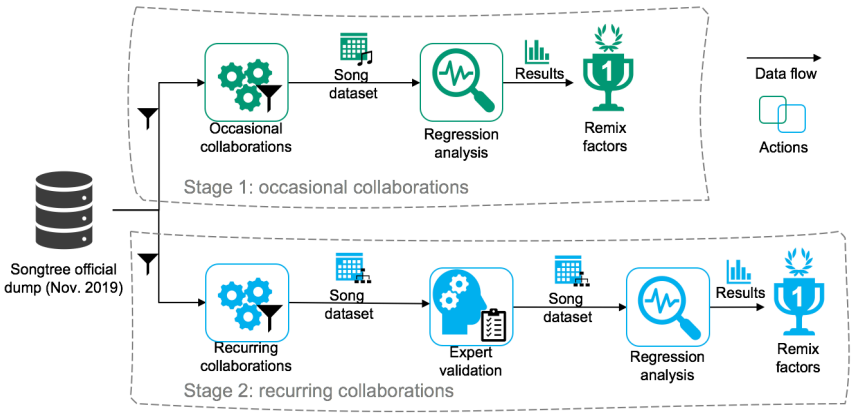

Figure 3. Workflows of the two stages of analysis performed in the study.

**4.1.2 Extension of Prior Work**

In this study, building on the evidence from related work on collaboration in online creative communities presented in Sect. 3.1, we extend our prior work on remix factors in Songtree by adding three new hypotheses. In addition, to carry out this study we build a new, larger dataset to fit a couple of more sophisticated regression models, as detailed next in Sect. 5.

In Songtree, authors can tag their songs with technical labels, for instance, to advertise the measure of the tempo in beats per minute (bpm) and the music key. Prior work on Wikipedia (e.g., [1,30,45,54]) has consistently found evidence that article metadata are reliable proxies for article quality and, as such, they can be used as antecedents of not-reverted page edits. Similarly, Luther et al. [35] found that Newgrounds animations that advertise their technical specifications are associated with higher chances of success. Building on these findings, we argue that songs tagged with technical metadata are more likely to be reused as well as receive more overdubs.

$\overline{H}6$: *Applying technical metadata tags to songs is positively associated with receiving overdubs.*

In addition, in Songtree users can interact indirectly by sending requests to overdub songs. Therefore, we speculate that songs for which overdub invitations have been sent are more likely to be overdubbed and receive more overdubs.

$\overline{H}7$: *Sending overdub invitations is positively associated with receiving overdubs.*

*Finally, Songtree allows members to interact also through the exchange of direct messages. Prior research building upon the common bond theory [37,44] has found that frequent communication is key to creating and maintaining strong online relationships. Therefore, we hypothesize that users are more likely to overdub songs by authors with whom they are in contact.*

$\overline{H}8$: *The amount of communication exchanged between two authors is positively associated with exchanging overdubs.*

*4.2 Stage 2: Recurring Collaborations*

In the second research stage, we undertake an exploratory analysis of recurring collaborations and uncover the antecedents of reuse for the songs composed through them. In the absence of any empirical evidence or general agreement in the literature (see Sect. 3.2), it is not possible to define hypotheses as we did for the first stage. Therefore, we refine the research question as follows.

First, we are interested in understanding how common recurring collaborations are in music co-creation communities like Songtree. Accordingly, we ask:

RQ1: Are there any instances of recurring collaborations in online music co-creation?

Having established the presence of recurring collaborations, we take interest in understanding their characteristics, in particular (i) the difference in community ranking between their members, (ii) their preferred genres and instruments, and (iii) the complementarity of their musical skills (e.g., instruments played). Therefore, we ask:

RQ2: What are the characteristics of the identified recurring collaborations?

Finally, having established in the first research stage the antecedents of reuse for songs generated through occasional collaborations, in the second stage we want to uncover the presence of potential differences as compared to songs written by members who collaborate frequently. Hence, we ask:

RQ3: What are the remix factors of songs generated through recurring collaborations?

## 5 EMPIRICAL STUDY

*5.1 First Stage: Analysis of Occasional Collaborations*

In this subsection, we provide a detailed description of the dataset (Section 5.1.1), the measures extracted (Section 5.1.2), and the regression-model selection strategy (Section 5.1.3).

**5.1.1 Dataset**

As the first step, we collected and analyzed the data extracted from the entire database dump of Songtree (from Nov. 2011 to Nov. 2019). The dataset was built from the entire dump of the database provided by the community administrators after signing a non-disclosure agreement. A breakdown of the data dump is reported in Table 1. As of November 2019, the community counted about 295K members, of which ~57K (20%) are authors who have recorded and shared at least one on Songtree; overall, over 263K songs had been uploaded to Songtree, of which ~200K new songs and ~63K overdubs.

As initial preprocessing steps, we first removed all the songs and their overdubs created before April 2015, i.e., when administrators were still participating actively to kick-start the community. Then, we filtered out all the new songs and their overdubs created in the last 27 days before the dump. This 27-day threshold corresponds to the 90[th] percentile of the overdub time intervals between the upload of a song and that of its first overdub. In other words, 90% of the songs in the rather sparse dataset have received their first overdub within 27 days since their upload. This step was necessary to avoid *right censoring* issues [3] and ensure that there was sufficient time to observe the event of interest (i.e., being overdubbed) for all the selected songs.

To build the final dataset, we further preprocessed the dump and excluded the content matching the following criteria:

Table 1. A comparison of data in the Songtree dump and the final dataset (rows in grey are filtered out).

|  | Original dump* | Final dataset** |
|---|---|---|
| **Users** | 295,193 | - |
| **Authors** | 57,868 | 49,517 |
| **Administrators** | 52 | - |
| **Songs** | 263,526 | 202,164 |
| **New songs** | 200,336 | 159,458 |
| **Overdubs** | 63,190 | 42,706 |
| **Contest songs** | 1,022 | - |
| **Closed songs** | 513 | - |
| **Hidden songs** | 47,504 | - |
| **Orphan songs** | 271 | - |
| **Self-overdubs** | 26,945 | - |
| **Remixes** | 1,199 | - |

*as of November 13, 2019; ** April 2105 – October 16, 2019*

- *Non-authors* – community members who have not recorded and shared any song on Songtree. They are excluded because they have gained no reputation as authors.
- *Administrators* – accounts registered by the members of the Songtree development team. We opted for excluding administrators' accounts and the content shared by them (e.g., contest songs) to avoid altering our findings on how the Songtree community behaves when collaborating.
- *Closed and complete songs* – songs that, as per the author's setting, either cannot be overdubbed or are marked as finished. Hence, they are excluded because, respectively, they disable or discourage overdubbing. Note that these are just leaf nodes. Instead, the song trees they belong to are retained.
- *Hidden songs* – songs that, as per the author's setting, are not publicly listed and can be found and overdubbed only if the author shares a link with others. These songs are used by authors who want their music to remain private or keep the collaboration restricted to their inner circle.
- *Orphan songs* – songs that belong to no song tree and overdubs derived from no parent songs.
- *Self-overdubs* – any child song derived from a parent song recorded by the same author. They are excluded because they do not represent meaningful cases of collaborative songwriting.
- *Remixes* – songs that only add effects or alter frequencies through the equalizer. They are excluded because they do not help their authors earn badges or improve their social ranking.
- *Contest songs* – songs uploaded by the Songtree team to start contests with prizes awarded to the best overdubs.
- *Recurring collaboration songs* – songs written collaboratively by frequent collaborators and analyzed in the second stage of the study.

At the end of the preprocessing stage, we ended up with a final dataset consisting of 202,164 songs (159,458 new songs + 42,706 overdubs), and 49,516 authors.

### 5.1.2 Measures

From the final dataset, we defined several measures to inform our analysis. In Table 2, we report each of the defined measures, along with its definition, scale (i.e., nominal, ordinal, interval, or ratio), and the hypothesis associated. Since the outcome measure for our statistical model is the number of overdubs received by a song, we define `#overdubs` as the dependent variable.

We note that the final dataset is cross-sectional, i.e., it provides a snapshot of multiple data collected at one point in time. Cross-sectional data are typically sensitive to reverse causality[5] problems. However, the availability of the official database dump (i.e., the entire history of events recorded in the community) enabled us to compute the value of the features at a point in time *just before* the event of interest '*song has been overdubbed.*' As an example, consider a song $S$ from our dataset, which has received $n>0$ overdubs. Let $O_S^x$ be any of its overdubs ($0<x\leq n$) and $T_S^x$ the time when $S$ received it. Accordingly, the experimental dataset contains a related record where the dependent variable to predict (i.e., the number of times $S$ was overdubbed) is $x$, and the measures regarding the song $S$ are calculated at the instant before $T_S^x$, that is, we compute the number of likes, plays, etc. received by $S$ *just before* it was overdubbed by $O_S^x$. In the rest of the paper, we refer to these features as *time-based*. For songs that received no overdubs, instead, all the measures are taken at the data collection time. As such, we can mitigate reverse causality issues and allow us to make inferences about the underlying direction of causality between the observed number of times a song has been reused and the occurrence of any of the predictors.

---

[5] Reverse causality refers to either a direction of cause-effect contrary to expectation or a two-way causal relationship between the predictors and the dependent variable.

Table 2. Measures grouped by level of analysis and hypothesis (the '*' indicates that a feature is not time-based). New hypotheses as compared to our prior work [7,8] are indicated with a bar (i.e., $\overline{H}$).

| Level | Measure | Scale | Description | H |
|---|---|---|---|---|
| Song | #overdubs | ratio | The number of overdubs received by the song | - |
| Song | #likes | ratio | No. of likes received by the song | H1 |
| Song | #bookmarks* | ratio | No. of times that the song has been bookmarked | |
| Song | #plays | ratio | No. of times the song has been played | |
| Song | #reposts | ratio | No. of times the song has been shared by other members in Songtree | |
| Song | #comments | ratio | No. of comments about the song entered in its comment thread | |
| Song | upload_time_interval | interval | Time difference (in minutes) between the respective upload times of an overdub (if any) and its parent song | H2 |
| Song | song_depth | ratio | The distance in number of nodes from the root song that started the song tree (0 for root songs) | H3 |
| Author | #followers | ratio | No. of users following author's activities on Songtree | H4 |
| Author | ranking | ratio | $\frac{\#followers + \#user\_likes + \#user\_plays + \#derived\_plays}{\#shared\_songs}$ | |
| Author | new_songs_badge | ordinal | Badge gained by uploading new songs. Values: {*None, Rookie, Songwriter, Composer*} | |
| Author | overdubs_badge | ordinal | Badge gained by overdubbing other authors' songs Values: {*None, Performer, Top performer, Virtuoso*} | |
| Author | overdubs_received_badge | ordinal | Badge awarded when enough overdubs are recorded an authors' songs. Values: {*None, Songsmith, Band leader, Maestro*} | |
| Author | has_avatar* | nominal | Whether the author has uploaded a profile picture or not. Values: {*Yes, No*} | H5 |
| Song | has_tags* | nominal | Whether the author applied any tags to the uploaded songs, such as the tempo in beats per minute (e.g., 4/4), the music key (e.g., Cmaj, G#), the instruments played and/or wanted (e.g., vocals, cello). Values: {*Yes, No*} | $\overline{H}6$ |
| Author | #invitations* | ratio | The numbers of invitations sent to other members to request an overdub of the song | $\overline{H}7$ |
| Author | msg_exchange_rate | ratio | The rate of messages exchanged between the author of the song and the author of its overdub | $\overline{H}8$ |

Following the work of Cheliotis et al. [11], the measures are described next according to two levels of analysis, namely *songs* and *authors*. For the song-related measures, we capture various dimensions and metadata as well as signals of appreciation for songs expressed by community members. We point out that #bookmarks and #invitations are not computed as time-based because the original database dump lacks the necessary piece of information. Instead, the dichotomous predictor related to metadata tags is not time-based because it is meant to capture specific static properties of songs.

Regarding the author-related measures, we capture the level of interactions between users as well as various signals and dimensions of the extent of their interactions, social ranking, productivity, and identity within the community. Below, we describe only those author-level features that do not have self-explanatory names, and for which reading the description in Table 2 is not sufficient. The measure ranking represents a coolness index updated weekly and used by Songtree administrators to rank the community members by status. It is computed per author as indicated in the formula in Table 2 where `#user_likes` is the cumulative number of likes received by all the songs by the author, `#user_plays` is the cumulative number of times that all the songs by the author have been played, `#derived_plays` is the cumulative number of times that all the songs derived from the author's songs have been played, and `#shared_songs` is the number of tracks uploaded by the author on Songtree. Regarding badges, the description of the values and thresholds to earn each of them is available online on the website.[6] Also, we note that `has_avatar` could not be extracted as time-based because the database dump does not contain any information regarding the time when authors add or change their profile pictures.

In Appendix A, we report the descriptive statistics of the measures. Regarding the song-level measures, we note a small mean value for the measures `#bookmarks` (1.4), `#reposts` (0.93), and `#comments` (1.34) along with small standard deviations, suggesting that these features are not very used in Songtree, unlike `#likes` (mean=7.46, SD=63.4) and `#plays` (mean=230, SD=2,355.25). The statistics regarding `upload_time_interval` show that the values for the measure are quite spread out (SD=29,492,939), with songs that were reused a mere 2 minutes (min) after and others that were reused after more than 4 years (max); on average, a reused song receives an overdub about 12.5 hours since their upload (mean=738,714 min.). The statistics of the measure `song_depth` reveal that most songs in the dataset are root songs (mean=0.36, SD=0.94). Regarding the use of tags, most songs in the dataset (213,368) have at least one tag. We also observe that 10 requests to overdub (`#invitations`) are sent on average for each song.

Concerning the author-level measures, we observe that the average rate of messages exchanged between a dyad of co-authors (`msg_exchange_rate`) is 65, albeit the measure is quite spread out (SD=2,643.77). Similar observation can be made for `#followers` (mean=56.1, SD=128.78) and `ranking` (mean=47.55, SD=293.15). Regarding the badges, we note that most users have not earned one. Finally, we note that most users have customized the avatar picture on their profile page.

**5.1.3 Regression-Model Selection Strategy**

Using the song- and author-level measures defined above, we build a regression model that predicts the number of overdubs received by songs. Because the dependent variable `#overdubs` can only take positive integer values (i.e., $\geq 0$), we perform a count data regression analysis, which is better suited to handle datasets with non-negative observations.

Different count data models can be used, whose choice depends on the characteristics of the data. In modeling count data, we follow the approach suggested by Cameron and Trivedi [10] and Green [25]. Two-part models are especially popular for modeling count data. In our specific case, to model the number of overdubs received by songs there is one part (binary or logistic), to determine whether a song is remixed at all, and a second part (count), to determine the consequent number of overdubs received for those with at least one overdub. Accordingly, in a two-part regression model, the count part contains the coefficients for the factor change in the expected count for those in the 'Not Always Zero' group (i.e., the songs that received one overdub or more) whereas the binary part contains the coefficients for the factor change in the odds of being in the 'Always Zero' group (i.e., the songs that received no overdubs) compared with the 'Not Always Zero' group [32].

To select the most appropriate two-part regression model, we compare the fit between a zero-inflated negative binomial (ZINB) model and a hurdle model. Albeit the selection of one model over the other can yield different results with different interpretations (e.g., [29]), in our study we show that, regardless of the theoretical speculations about which is more appropriate, they lead to consistent conclusions. For further details on the regression model selection strategy, please refer to Appendix B.

---

[6] https://songtr.ee/badges.php

*5.2 Second Stage: Analysis of Recurring Collaborations*

For the second stage of analysis, due to the exploratory nature of our analysis, we follow a mixed-methods approach characterized by a sequential explanatory strategy [17]. First, we use data mining techniques to extract recurring collaborations from the dataset; we also repeat the regression analysis described in the first stage to uncover remix factors in recurring collaborations. Then, we consult the Songtree founder (also CEO and lead developer) to garner further insights on recurring collaborations and assess the results of our mining analysis against his understanding of the phenomenon of recurring collaborations. We follow an expert validation approach, which is often used in social and medical science (e.g., [18]) where researchers consult with qualified experts in the field of interest to collect judgments, informed opinions, and assessments through surveys and interviews.

In the rest of this subsection, we first provide some background on the data mining techniques used to mine recurring collaborations (Section 5.2.1) and how they were adapted to the music domain (Section 5.2.2). Then, we illustrate the dataset built for the frequent pattern mining analysis (Section 5.2.3) and the expert validation interview protocol (Section 5.2.4).

**5.2.1 Frequent Pattern Mining**

The classical problem of association pattern mining [2] is typically defined in the context of supermarket data containing sets of items bought by customers, which are referred to as *transactions*. The goal is to determine associations between groups of items bought together, which can intuitively be viewed as a *k*-way correlation between items. The most popular model for association pattern mining uses the frequencies of sets of items as the quantification of the level of association. The discovered sets of items are referred to as *frequent itemsets* or *frequent patterns*.

Frequent itemsets are used to generate *association rules* in the form $X \Rightarrow Y$, where $X$ and $Y$ are sets of items. Association rules are intended to capture dependencies among items in a dataset. For example, for the itemset {Eggs, Milk, Yogurt}, the association rule {Eggs, Milk}⇒{Yogurt} suggests that buying eggs and milk makes it more likely to also buy yogurt.

Let $T$ be a dataset containing a set of n transactions, denoted by $T_1 \ldots T_n$. An *itemset* $X=\{i_1,\ldots,i_k\}$ is a set of items and a *k-itemset* denotes a set of items of cardinality $k$. The fraction of transactions in $T_1 \ldots T_n$ in which an itemset occurs as a subset provides a quantification of its frequency known as *support* and denoted with $\sup(X)$. In other words, $\sup(X)$ is the measure of how frequently the collection of items $X$ occur together as a percentage of all transactions.

The goal of frequent pattern mining techniques is to identify frequent itemsets whose support is larger than a chosen threshold (*minsup*). For example, the association rule $X \Rightarrow Y$ is selected if $X$ and $Y$ occur together in at least *minsup*% of the $n$ total transactions in the dataset, i.e., when $\sup(X \Rightarrow Y) = \frac{\text{transactions containing } X \cup Y}{n} > minsup$.

The minimum-support criterion ensures that enough transactions are relevant to the rule and, therefore, it has the required critical mass to be considered relevant to the application at hand. Yet, we need a second criterion to also ensure that the rule has sufficient strength in terms of conditional probability, i.e., that the antecedent and consequent in the rule are dependent.

To establish whether a dependence exists between the items in a rule, Brin et al. [5] proposed to use correlation-based measures that are more suited in domains other than market data mining. Consistently, for the second criterion, we use the *lift* (or *interest*). The lift of a generic rule $X \Rightarrow Y$, computed as in equation (1), tells us how likely it is for the items in $Y$ to be purchased when the items in $X$ are purchased while controlling for how popular the items in $X$ and $Y$ are in the dataset.

$$lift(X \Rightarrow Y) = \frac{\sup(X \cup Y)}{\sup(X) \cdot \sup(Y)} \quad (1)$$

The statistical definition of independence between two generic events $A$ and $B$ is that the probability of $A$ and $B$ occurring together equals the product of their *apriori* probability (i.e., $P(A \wedge B)/P(A) \cdot P(B) = 1$).[7] Therefore, the further from 1 the lift of an association rule is, the more dependent its antecedent and consequent are. In other words, greater lift values indicate stronger associations between items and increase the confidence that their co-occurrences in transactions are not spurious.

**5.2.2 Frequent Pattern Mining Algorithm and Rules Generation**

When an itemset *I* is contained in a transaction *T*, all its subsets will also be contained in the transaction. Therefore, the support of any subset *J* of *I* will always be at least equal to that of *I*. This property is referred to as the *monotonicity property*. The monotonicity property of support implies that every subset of a frequent itemset will also be frequent. This is referred to as the *downward closure property*. One of the most well-known algorithms for frequent pattern mining is *Apriori*, which builds on this heuristic to prune the search space. The algorithm uses an iterative approach in which the frequently identified *k*-itemsets are exploited to search for (*k+1*)-itemsets. Then, based on the downward closure property, if a *k*-itemset is not frequent, all its (*k+1*)-super-patterns will not be frequent and can be pruned. In other words, if the support of an itemset does not exceed the predetermined threshold (*minsup*), according to which an itemset is to be considered frequent, then this will also apply to all supersets that contain it.

In Appendix C, we provide a detailed description of the pattern mining algorithm adapted to extract frequent collaboration as well as the procedure to generate association rules in the form {*Author1*,…, *AuthorN-1*}⇒{*AuthorN*} from the *k*-itemsets.

**5.2.3 Dataset**

The dataset for the analysis of recurring collaborations was built similarly to the first stage. From the original dump containing 367,421 trees and 542,140 nodes, we filtered again the orphaned nodes lacking a reference to a tree, the contest trees, which foster occasional collaborations, and all trees containing only the root node, which do not entail any collaboration. Eventually, we obtained a filtered subset containing 21,702 trees, 84,613 nodes, and 11,000 unique song authors, which was fed to the frequent pattern mining algorithm.

To mine the recurring collaborations (i.e., the overdub chains extracted from traversing a tree from its root to the leaves) and identify the frequent itemsets of recurring collaborations' members, we transformed the subset from a tabular format into a more appropriate data structure based on nested linked lists. Thus, we created a linked list for each tree therein. Each of these lists contains in turn a list for all the paths obtained traversing the trees; finally, for each path, we created a list containing the usernames of the song authors.

Eventually, we identified 47 unique authors who are involved in recurring collaborations, which generated 2,141 songs (more details are provided in Section 6.2). We notice that these songs have been filtered out from the previous dataset of occasional collaborations described in Section 5.1.1.

**5.2.4 Expert Validation**

We conducted a one-hour interview with the Songtree founder on Oct. 2020 over Zoom. Albeit we prepared a shortlist of predetermined questions, the interview unfolded intentionally in a semi-structured, conversational manner to offer the chance of exploring issues in follow-up questions. First, we asked a couple of questions about recurring collaborations in Songtree, to understand whether he agreed with the definition and to what extent he believed the phenomenon was present in the community. Then, we presented the list of the recurring collaborations identified and asked whether he would consider the members as frequent collaborators according to our conceptualization, his experience, and data. Finally, we asked him to provide examples, if any, of recurring collaborations that we had missed.

---

[7] The generalization to more than two events is: $P(A \wedge ... \wedge Z)/P(A) \cdot ... \cdot P(Z) = 1$

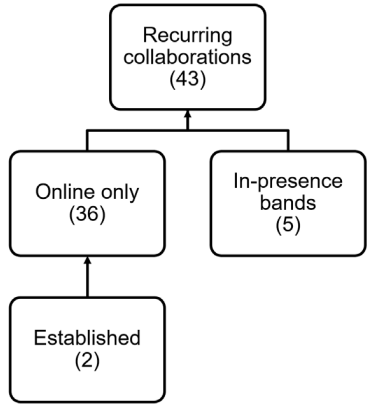

Figure 4. The taxonomy of recurring collaborations. The numbers between parentheses are the instances identified.

## 6 RESULTS

In this section, we present the results of our study. First, in Sect. 6.1 we present the findings regarding the presence and types of recurring collaborations in Songtree. Then, in Sect. 6.2 we report the results of the respective count data regression analysis to compare the remix factors in occasional and recurring collaborations.

### 6.1 Identification of Recurring Collaborations

In this section, we present the recurring collaborations identified in Songtree. Figure 4 illustrates the observed taxonomy of recurring collaboration types. They can be *online-only*, i.e., members' interaction happens exclusively online, mediated by the Songtree platform. *Established* collaborations are a subtype of online-only collaborations whose participants team up, choose a name, and explicitly acknowledge their affiliation in their username or profile page. Finally, the last type identified is that of *in-presence* collaborations, i.e., actual bands that primarily exist offline and make live music but also share content on Songtree.

**Online-only**. After executing the frequent pattern mining algorithm reported in Appendix C, we created a dictionary containing all the $k$-itemsets and their occurrences. After filtering out those occurring < 3 times (assumed as infrequent), we obtained 1,165 2-itemsets, 102 3-itemsets, and 9 4-itemsets. By dividing the number of occurrences of each $k$-itemset by the total number of transactions in the dataset, we obtained the support for each $k$-itemset, necessary to compute the lift score of the generated association rules (see Sect. 5.2.2). After applying the lift > 1 filter for retaining the positive associations only, we were left with the following candidate $k$-itemsets: 579 2-itemsets and 22 3-itemsets. Regarding the *minsup*, we chose not to apply any filter. Given the exploratory nature of frequent pattern mining analysis applied to online music collaboration, we did not have at our disposal any reference to inform the choice of the *minsup* value. Hence, rather than choosing arbitrarily a threshold, we opted for not applying any filter at this stage and, instead, established a value in retrospection.

While applying the lift filter ensures that we select authors who are associated based on their production activity, one still may argue that being in a band involves some sort of interaction and socialization. Therefore, we analyzed the degree of interaction quantified as the number of messages and invites exchanged between authors, motivated by the evidence reported by Stanko [50] and Luther et al. [34] that successful collaborations are associated with frequent communication and previous interactions. Furthermore, as observed by Settles and Dow [46], members of recurring collaborations tend to have a similar ranking in the community; accordingly, we also analyzed the differences (delta) in their coolness index and the number of likes received.[8]

Table 3. Results of the *k*-means clustering. The grey background indicates the bins filtered out after applying the filter for removing infrequent collaborations.

| Category | Low | Medium | High |
| --- | --- | --- | --- |
| # Messages | [0, 197] | - | [198, 450] |
| # Invites | [0, 89] | [90, 279] | [280, 446] |
| |Δ Likes| | [0, 12] | [13, 36] | [37, 61] |
| |Δ Coolness index| | [0, 26] | [27, 106] | [107, 168] |

Because all these features (# message, # invities, Δ likes, and Δ coolness index) are continuous and present a long-tail distribution, we used discretization to transform them into ordinal features and make sense. We first computed the optimal number of clusters; then, we applied the *k*-means clustering algorithm to group the authors according to each feature category. We identified three clusters (called *low*, *medium*, and *high*) for all the features. Details on the clustering are reported in Table 3. Finally, we proceeded with filtering all the *k*-itemsets with < 3 collaborations, which left us with 36 candidates for online-only recurring collaborations (the complete list is available in Appendix D). We observe that among these instances there are no cases of a *high/medium* difference in the number of likes and coolness index among the members – i.e., all have a similar ranking in the community – as well as a *low* number of messages and invites between their members. These observations are consistent with previous evidence reported in Sect. 3.1, thus increasing the confidence that the instances of recurring collaborations identified are not spurious.

All the online-only collaborations except one are formed by two members and count 47 unique authors overall, of which 32 participate in only one collaboration and the remaining 15 participate in two to six collaborations. The most recurring genre is rock (29 songs), followed by hip-hop and acoustic (28), alternative (21), electronic, and rap (19). Regarding instruments, the most recurring ones are voice (28), followed by acoustic guitar and drums (15).

Next, we analyzed the dataset to identify any cases of established online-only collaborations whose members explicitly acknowledge their affiliation. To identify such cases, we used the following semi-automated approach, consisting of three steps. First, we hypothesized that members might name these collaborations and use such names as part of their username in Songtree (e.g., *Queen/Freddie*, *Queen–Brian*). To perform this analysis, we computed the Levenshtein distance [58], a measure of similarity between two strings, and then applied sequence matching. Second, we analyzed the text reported in the bio section of the profile page of community members. We marked as candidates those profiles that use expressions such as '*trio*', '*quartet*', '*band*', '*collaboration*', and '*together*' or pronouns like '*we*', '*us*', and '*our*.' Third, we hypothesized that established bands may share the same account so that collaboration happens by uploading overdubs in turn to extend the existing track (e.g., first the guitar base, then bass, drums, and finally voice); as such, we re-scanned the same data structure used to extract transactions to identify cases of recurring self-overdubs, i.e., paths in song trees where the same user uploads three or more consecutive overdubs. After executing these three steps, we obtained an overall set of 220 user profiles, two of which were confirmed to be actual instances of established, online-only recurring collaborations after manual inspection (see Table 4). The first instance, *Sludge*, is a collaboration involving 2 to 6 members (judging from the instruments played) who play grunge/nu metal music. Regarding the other instance, *Vinnie & the Poets*, their profile page contains a link to the nominal singer's website, but no information about the other members. Furthermore, the band *Vinnie & the Poets* provides an interesting case. Together with the user *Raul* (an active member with over 1,000 uploads and 44 coolness index), they form the only three-member recurring collaborations identified, with 11 songs where user *Raul* always contributes by playing piano, thus highlighting that *Vinnie & the Poets* are lacking such skill.

---

[8] We initially considered including also the number of plays, but it turned out to be strongly correlated with that of likes received.

Table 4. The two instances of established online-only collaborations (the names are fictional values to preserve privacy).

| Band | Description | #self-overdubs | #songs | Instrument | #songs per instrument | Date (yyyy-mm) First | Date (yyyy-mm) Last |
|---|---|---|---|---|---|---|---|
| Sludge | French Grunge / Nu Metal Band | 7 | 25 | Agida | 6 | 2019-01 | 2019-11 |
| | | | | Drums | 17 | | |
| | | | | Guitar (acoustic) | 5 | | |
| | | | | Guitar (electric) | 23 | | |
| | | | | Piano | 1 | | |
| | | | | Voice | 14 | | |
| Vinnie & the Poets | Native American Latin Jazz | 7 | 52 | Bass | 7 | 2016-07 | 2017-12 |
| | | | | Bass (electric) | 2 | | |
| | | | | Flute | 4 | | |
| | | | | Guitar (acoustic) | 22 | | |
| | | | | Guitar (electric) | 28 | | |

**In-presence bands**. Here we sought instances of in-presence collaborations, that is, bands who also make and play music outside of the Songtree platform. Accordingly, we performed an analysis of usernames and profile pages like the one carried out to identify established, online-only collaborations, based on the same assumption that bands must have a name.

The analysis returned 7 candidates of which 2 had already been confirmed as established online-only collaborations and, therefore, were excluded. The manual inspection confirmed that the remaining 5 instances are in-presence bands (see Table 5). Overall, we notice that they are not very active in the community, and a couple of them (*Priest Dream* and *Nope*) have been active only for a few months.

The in-presence band with the highest coolness index (6) is *Palace*, with 47 songs shared between February 2016 and June 2019. The profile page does not provide details or external links; hence, we ignore how many members are in the band. However, the analysis of songs uploaded shows some tracks labeled as *Multiple instruments* (8 out of 46), which we confirmed to be full songs released by the band rather than incremental tracks. This observation is in line with our speculation that in-presence band accounts may be used by one member. Another relevant aspect of this band is the high number of invites sent by the band (more than 300), some to user *Bulls* (a moderately influential member of Songtree, with coolness index 15), who has extended some of the songs by adding another voice. While the association between them is not strong enough to consider them an online-only recurring collaboration, it highlights that the *Palace* band arguably looks at Songtree as an opportunity to get further visibility as well as to compensate for the lack of another singer.

Table 5. The five in-presence bands identified (the names fictional to preserve privacy).

| Band | Mean #likes | Band coolness index | #songs | Date First | Date Last |
|---|---|---|---|---|---|
| 3M and EastMusic | 4 | 0 | 26 | 2017-05 | 2018-10 |
| Priest Dream | 5 | 1 | 15 | 2015-11 | 2016-03 |
| Palace | 11 | 6 | 46 | 2016-02 | 2019-06 |
| Kate & Kun | 1 | 0 | 9 | 2016-11 | 2017-06 |
| Nope | 8 | 1 | 24 | 2018-05 | 2018-07 |

**Instruments and genres.** Next, we take interest in understanding the role played by music genres preference and skills in recurring collaborations; specifically, we compared whether users collaborating recurrently play the same or different instruments in occasional collaborations and likewise for music genres. Accordingly, first, for each member of the recurring collaborations identified we generated two disjoint sets of songs, one for songs generated as a result of a frequent collaboration and the other for those resulting from occasional collaborations. Then, for each of these two sets of songs, we generated two sets of tags, one for the genres and the other for the instruments played in the songs. Finally, we computed the Jaccard similarity index [27], a statistic used for gauging the similarity of sample sets within the range [0,1]. Regarding the similarity of genres, we find that on average (median=0.43, SD=0.16) there is a ~40% similarity among the genres played in occasional collaboration as compared to recurring collaborations. The Jaccard index suggests that occasional collaborations may be used to explore and experiment with diverse types of music. Consistently, the median number of different genres played in recurring collaborations is 4.5 whereas it is 11 in occasional collaborations. Regarding the instruments played, the median Jaccard similarity index is 0.5 (SD=0.28), showing that Songtree members do not always play the same instruments in both recurring and occasional collaborations.

**Expert validation**. During the interview with the Songtree founder, we presented the list of recurring collaborations identified, separated by type. Regarding the 2 established, online-only collaborations and the 5 in-presence bands, we browsed together the profiles of each member and verified that they did acknowledge the affiliation. He then provided us with a couple of further instances of collaborations that he remembered being quite active in the past and even surveyed internally before. We realized that we had failed to properly identify them as established because of the lack of any references to the recurring collaboration in their usernames and profile page.

Regarding the 36 instances of online-only collaborations, the Songtree founder flagged 10 'suspicious' instances–involving members who did not strike him as recurrent collaborators–and manually inspected them by looking up the track record of each member. He was then able to confirm that the suspicious cases did have a good number of overdubs together over the total number of overdubs uploaded. Using the same approach for the remaining cases, he confirmed that there were no cases of spurious collaborations identified as online-only recurring collaborations (i.e., no false positives). Finally, he gave us five examples of long-time, very active Songtree users who he expected to find involved in recurring collaborations but were not on the list. We were able to confirm that they were not included because of how the lift formula (1) in Section 5.2.1 is defined. Uploading more songs implies a larger denominator in the formula (i.e., increased support). In other words, large productivity increases the chances for users to have occasional collaborations but, at the same time, raises the bar to consider them recurring.

Finally, in retrospection, we observe that the smallest *minsup* value associated with the mined patterns of frequent collaborations is 0.005, with a maximum of 0.037 (average 0.012).

---

**Recurring collaborations in Songtree.** We identify 43 instances of recurring collaborations, of which 36 are online only, 2 are established, and 5 are in-presence bands. In retrospection, the *minsup* threshold (smallest) value associated with the mined patterns of recurring collaborations is 0.005.

**Characteristics of recurring collaborations.** All recurring collaborations involve two or three members with a very similar ranking in the community. The most recurring genres played are rock and hip-hop. Regarding instruments, the most frequent are voice, acoustic guitar, and drums. Regarding the similarity between occasional and recurring collaborations, we find a 40% similarity in terms of genres played and 50% in terms of instruments played.

*6.2 Analysis of Remix Factors: Occasional vs. Recurring Collaborations*

In this section, we describe the count data regressions modeling the associations of the song- and author-related remix factors with the dependent variable `#overdubs`. In addition, we compare the results for occasional and recurring collaborations.

**6.2.1 Data preparation**

To build the count data models, we used several functions as implemented in R ver. 4.1.0. Concerning recurring collaborations, we focused on analyzing the N=2,141 songs written by the 38 instances of online-only collaborations identified. We specifically chose online-only collaborations because, other than being more common, their activity is consistent with the Songtree community's goal, i.e., to provide a platform for online music co-creation. Furthermore, for in-presence bands, it is likely the existence of other real-life factors (e.g., familiarity, physical interaction), which would be hardly controllable in our analysis. Regarding occasional collaborations, we point out the number of instances in the R dataset is N=218,902, which is more than the N=202,164 songs retained in the final dataset obtained after preprocessing (see Table 1). This discrepancy depends on the internal representation used to serialize a song tree. For example, consider a song tree with 3 nodes, the root song A and its two overdubs B and C. This tree is serialized as 4 entries in the R dataset, namely i) *A overdubbed by B*, ii) *A overdubbed by C*, iii) *B not overdubbed*, and iv) *C not overdubbed*.

All the continuous variables in both datasets follow a long-tail distribution and, therefore, were log-transformed to reduce skewness. Furthermore, as multicollinearity reduces inferential ability by affecting standard error estimates, we checked our dataset for multicollinearity problems [39]. Accordingly, we computed the correlation matrices. In the case of recurring collaborations, we found that the `#likes` predictor has a strong correlation with `#plays`, `#reposts`, and `#comments`, so we removed it. Also, `ranking` is strongly correlated with `#follower`, which was therefore discarded. Finally, `msg_exchange_rate` is strongly correlated with `upload_time_interval`; to fix the problem, we retained `upload_time_interval` and replace `msg_exchange_rate` with `#sent_messages` and `#received_messages`. In the case of occasional collaborations, because the predictors `#plays` and `ranking` have a strong pairwise correlation (i.e., $\geq 0.7$), we retained the latter and discarded the former.

**6.2.2 Model selection**

Before delving into the presentation of results of the count-data model regression analyses, we briefly comment on the model selection process (see Appendix B for more).

In the case of the recurring collaborations, the Likelihood-Ratio Test of overdispersion is significant ($\chi^2$=2,933, p < 0.001), which leads us to reject the null hypothesis of equidispersion and, therefore, use the negative binomial distribution instead of Poisson to develop a count data model. Accordingly, we fit both the zero-inflated negative binomial (ZINB) and the hurdle models. We find that the former fits the data better (AIC 5,594 and 5,947, respectively) and that this difference is significant according to the Vuong test (z=7.658, p<.001). Hence, in the following we only report (see Table 6) and discuss the results obtained with the hurdle model; nonetheless, we underline that the conclusions would not change if the hurdle model was used instead.

Regarding occasional collaborations, because the Likelihood-Ratio Test of overdispersion is also significant ($\chi^2$=29,721, p < 0.001), we built both the ZINB and the hurdle models fit using the model selection strategy illustrated before. However, for the sake of brevity, we only report (see Table 7) and discuss the ZINB model because it provides a better fit than the hurdle model (AIC 121,994 and 129,793, respectively) and the difference is statistically significant according to the Vuong test (z=37.225, p<.001). Also in this case, using either model leads to the same conclusions.

**6.2.3 Results**

In the following, we report and compare the results of the two regressions performed on the datasets of songs created through recurring (Table 6) and occasional collaborations (Table 7). Along with the estimated β coefficients, we also report in each table the factor change coefficients exp(β). Specifically, for the binary part of each regression model, the exp(β) coefficient of an explanatory variable *X* corresponds to the odds ratio,[9] i.e., the expected change in odds for a log-unit increase in the explanatory variable *X*; for the count part, the exp(β) coefficient is the change in the expected count of `#overdubs` for a log-unit increase in the explanatory variable *X*. The interpretation of the coefficients, and their relative importance in the case of hypotheses operationalized using multiple variables, is clarified by the following examples taken from Table 6. One log-unit increase in `#plays` of a song is significantly and positively is associated with an expected count of `#overdubs` received that is 48% higher (exp(β)=1.48), holding the other variables constant; as such the variable `#plays` prevails over `#bookmarks`, which is associated with an expected count of `#overdubs` received that is 19% higher (exp(β)=1.19). Regarding the variable `song_depth`, one log-unit increase in the depth of the song within the tree is negatively associated with a 57% decrease in the odds of a song being overdubbed at all (exp(β)=0.63), holding the other variables constant.

Table 6. Factor change β coefficients of the Hurdle model for the number of overdubs received by songs created through recurring collaborations. In the count part, exp(β) is the change in the expected count for a unit increase in predictor *X*; in the binary part, exp(β) is the change in odds for a unit increase in *X*. Significant predictors are shown in **bold** (sig.: \*\*\*p<0.001; \*\*p<0.01; \*p<0.05).

| Level (Song / Author) | Hypothesis | Predictor *X* | Dependent variable: #overdubs, N=2,141 | | | | | |
|---|---|---|---|---|---|---|---|---|
| | | | Count equation | | | Binary equation | | |
| | | | β coeff. | exp(β) | Clustered SE | β coeff. | exp(β) | Clustered SE |
| S | H1 | #plays | **0.389\*\*\*** | **1.48** | **007** | **0.476\*\*** | **1.61** | **0.18** |
| S | | #bookmarks | **0.173\*\*\*** | **1.19** | **0.08** | **0.182\*\*** | **1.20** | **0.22** |
| S | | #reposts | -0.051 | 0.95 | 0.07 | -0.049 | 0.95 | 0.16 |
| S | | #comments | -0.070 | 0.93 | 0.04 | -0.250 | 0.78 | 0.14 |
| S | H2 | upload_time_interval | -0.036 | 0.96 | 0.02 | **-0.142\*\*\*** | **0.87** | **0.03** |
| S | H3 | song_depth | **-0.469\*\*\*** | **0.63** | **0.11** | **-1.028\*\*\*** | **0.36** | **0.28** |
| A | H4 | ranking | -0.139 | 0.87 | 0.12 | **0.627\*\*\*** | **1.87** | **0.29** |
| A | | new_songs_badge=Rookie | 0.035 | 1.04 | 0.28 | **-0.022\*** | **0.98** | **0.25** |
| A | | new_songs_badge=Songwriter | **-0.135\*** | **0.87** | **0.16** | 0.151 | 1.16 | 0.35 |
| A | | new_songs_badge=Composer | 0.283 | 1.33 | 0.26 | **-0.305\*** | **0.74** | **0.29** |
| A | | overdubs_badge=Performer | 0.047 | 1.05 | 0.25 | 0.016 | 1.02 | 0.31 |
| A | | overdubs_badge=Top_performer | 0.109 | 1.16 | 0.24 | **-0.125\*** | **0.88** | **0.33** |
| A | | overdubs_badge=Virtuoso | **-0.251\*** | **0.78** | **0.21** | **-0.324\*\*** | **0.72** | **0.35** |
| A | | overdubs_received_badge=Songsmith | 0.498 | 0.61 | 0.29 | **0.230\*\*\*** | **1.26** | **0.33** |
| A | | overdubs_received_badge=Maestro | **0.349\*\*** | **1.42** | **0.24** | 0.158 | 1.17 | 0.35 |
| S | H̄6 | has_tags=True | 0.105 | 1.11 | 0.23 | 0.636 | 1.89 | 0.51 |
| S | H̄7 | #invitations | 0.001 | 1.00 | 0.03 | **0.195\*\*\*** | **1.22** | **0.07** |
| A | H̄8 | #sent_messages | **0.026\*** | **1.03** | **0.08** | **0.112\*** | **1.20** | **0.16** |
| A | | #received_messages | **0.029\*** | **1.03** | **0.11** | **0.291\*** | **1.34** | **0.18** |
| | | Log Likelihood | -2,748 | | | | | |
| | | Vuong test (z) | 7.658\*\*\* | | | | | |
| | | AIC | 5,594 | | | | | |

In addition, because the datasets contain repeated observations (i.e., multiple songs uploaded by the same author or collaboration), in Tables 6 and 7 we report the clustered Standard Errors (SE) to deal with the unmet assumption of independence in regression models. In the case of occasional collaborations, standard errors were computed at the subject level whereas they were clustered by collaboration (i.e., dyad or triad) in the case of recurring collaborations.

---

[9] An odds ratio is a measure of association between an exposure and an outcome. It represents the odds that an outcome (in this case, a song receiving an overdub) will occur given a particular exposure (one of the considered predicting variables), compared to the odds of the outcome occurring in the absence of that exposure. Values > 1 indicate a positive association, those < 1 indicate a negative one.

Table 7. Factor change β coefficients of the ZINB model for the number of overdubs received by songs created through occasional collaborations. In the count part, exp(β) is the factor change in the expected count for a log-unit increase in predictor *X*; in the binary part, exp(β) is the factor change in odds for a log-unit increase in *X*. Significant predictors are shown in **bold** (sig.: ***p<0.001; **p<0.01).

| Level (Song / Author) | Hypothesis | Predictor *X* | Dependent variable: #overdubs, N=218,902 | | | | | |
|---|---|---|---|---|---|---|---|---|
| | | | *Count equation* | | | *Binary equation* | | |
| | | | *β coeff.* | *exp(β)* | *Clustered SE* | *β coeff.* | *exp(β)* | *Clustered SE* |
| S | H1 | #likes | **0.549*** | **1.73** | **0.08** | **1.864*** | **6.50** | **0.05** |
| S | | #bookmarks | **0.310*** | **1.36** | **0.04** | **0.628*** | **1.87** | **0.07** |
| S | | #reposts | **0.217*** | **1.24** | **0.05** | **0.976*** | **2.65** | **0.06** |
| S | | #comments | **-0.149*** | **0.86** | **0.04** | -0.394 | 0.67 | 0.21 |
| S | H2 | upload_time_interval | **0.258*** | **1.29** | **0.02** | **-0.759*** | **0.47** | **0.08** |
| S | H3 | song_depth | **-0.481*** | **0.62** | **0.12** | **-0.784*** | **0.46** | **0.04** |
| A | H4 | #followers | **0.114*** | **1.12** | **0.04** | **0.736*** | **2.09** | **0.05** |
| A | | ranking | **0.027*** | **1.03** | **0.04** | **0.973*** | **2.65** | **0.11** |
| A | | new_songs_badge=Rookie | **-0.380*** | **0.68** | **0.19** | -0.075 | 0.93 | 0.25 |
| A | | new_songs_badge=Songwriter | **-0.313*** | **0.73** | **0.21** | **-1.102**** | **0.33** | **0.30** |
| A | | new_songs_badge=Composer | **-0.399*** | **0.67** | **0.25** | **-0.865*** | **0.43** | **0.35** |
| A | | overdubs_badge=Performer | **-0.196*** | **0.82** | **0.08** | 0.117 | 1.12 | 0.37 |
| A | | overdubs_badge=Top_performer | 0.019 | 1.02 | 0.11 | **-0.935**** | **0.39** | **0.43** |
| A | | overdubs_badge=Virtuoso | **-0.328*** | **0.72** | **0.20** | **-0.943**** | **0.39** | **0.30** |
| A | | overdubs_received_badge=Songsmith | **0.751*** | **2.12** | **0.10** | **0.355*** | **1.43** | **0.11** |
| A | | overdubs_received_badge=Band_leader | **1.190*** | **3.28** | **0.20** | **0.497*** | **1.64** | **0.15** |
| A | | overdubs_received_badge=Maestro | **1.490*** | **4.42** | **0.25** | **0.720*** | **2.05** | **0.20** |
| A | H5 | has_avatar=True | **0.186*** | **1.20** | **0.17** | **0.160*** | **1.17** | **0.02** |
| S | $\overline{H6}$ | has_tags=True | -0.017 | 0.98 | 0.19 | **0.816*** | **2.26** | **0.02** |
| S | $\overline{H7}$ | #invitations | 0.032 | 1.03 | 0.02 | 0.059 | 0.94 | 0.09 |
| A | $\overline{H8}$ | msg_exchange_rate | -0.033 | 0.97 | 0.01 | -0.248 | 0.78 | 0.15 |
| | | *Log Likelihood* | -60,952 | | | | | |
| | | *Vuong test (z)* | 37.225*** | | | | | |
| | | *AIC* | 121,994 | | | | | |

Next, we comment on the results of our regression analyses for each hypothesis.

**Popular songs**. As regards *H1* (*the number of reactions generated by songs is positively associated with receiving overdubs*), we find partial confirmation. Among the overdubbed songs recorded in recurring collaborations (see Table 6), we observe that a log-unit increase in `#plays` and `#comments` of a song is associated respectively with a 48% and +19% increase in the expected count of overdubs received. In addition, a log-unit increase in the same predictors is associated respectively with a 61% and 20% increase in the odds of being overdubbed at least once, respectively. As compared to the previous findings on occasional collaborations (see Table 7), we find consistent results regarding the `#bookmarks` and `#plays` (highly correlated with the `#likes`). Therefore, one conclusion for the first hypothesis *H1* is that popular songs that receive positive feedback and appreciation have in general significantly higher odds of being overdubbed and receiving more overdubs. Instead, we find mixed results regarding the predictors `#comments` and `#reposts`, which are not significant for songs resulting from recurring collaborations whereas they have, respectively, a negative and positive association with the dependent variable in the case of occasional collaborations (see Table 7). We already noticed that these features are not common in the occasional collaborations dataset and even less so when it comes to recurring collaborations.

While we reserve to investigate them further, we speculate that these mixed findings may be related to the limited use of these features in the community – in the original dump, the songs have on average about one comment and one repost each.

**Recent and mature songs**. As regards *H2* (*time is negatively associated with receiving overdubs*), the results indicate partial support for the hypothesis. For recurring collaborations (see Table 6), we find that a log-unit increase in the `upload_time_interval` predictor is associated with a 13% decrease in the chances of songs being overdubbed at least once – i.e., songs that do not receive the first overdub soon after being uploaded will likely never be remixed at all – whereas the predictor shows no association with the expected count of overdubs. The results of the regression analysis on occasional collaborations show consistent results for the binary part, whereas the count part shows that for songs that have been reused once or more, the time since the upload is associated with a positive variation in the expected overdubs received (+29%, see Table 7) – i.e., they can attract more overdubs over time. Therefore, the comparison suggests that songs generated through recurring collaborations do not benefit from any cumulative advantage deriving from receiving more overdubs over time, an indication that if these songs are not are soon reused by recurring collaborators they likely will not be overdubbed at all.

Regarding *H3* (*the degree of derivativity of songs negatively associated with receiving overdubs*), for recurring songs (see Table 6) we find that a log-unit increase in `song_depth`, is associated with a 64% decrease in the odds of a song receiving at least one overdub, and a 37% decrease in the expected counts of `#overdubs` received. Similar findings can be observed in Table 7 for occasional collaborations (-54% and -38%, respectively). Overall, we notice that derived songs are always less generative, regardless of whether they are the results of occasional collaborations rather than recurring ones. As such, we find support for *H3* given that more mature songs created towards the end of a long collaboration process are less likely to be overdubbed and receive fewer overdubs.

**Reputation.** Regarding *H4* (*the ranking of authors in the community positively associated with receiving overdubs*), the results of our study provide mixed support for our hypothesis that the reputation of authors in the community is positively associated with a higher likelihood of their songs being overdubbed as well as a higher count of overdubs received. For recurring collaborations (see Table 6), we observe that `ranking` is not a significant predictor of song reuse. Conversely, for occasional collaborations (see Table 7) we find that a log-unit increase in both authors' `ranking` and `#followers` is associated with a significant increase in the odds of a song being reused (+109% and +165%, respectively) as well as the expected count of received overdubs (+12% and +3%). The lack of support for recurring collaborations is arguably explained by the small delta in `ranking` between the parties. As for the badges, the results are consistent for both types of collaborations since we find that `new_songs_badges` and `overdubs_badges` are associated with a decrease in the odds of songs being overdubbed as well as the expected count of overdubs received. Instead, `overdubs_received_badges` are consistently associated with an increase in the odds of a song being overdubbed at least once as well as the expected count of the overdubs received. We speculate that this is because the `overdubs_received_badges` gauge the extent to which one's songs are remixed by others and the reputation thus gained in the community, whereas the other two types of badges measure one's productivity in terms of songs and overdub uploaded regardless of whether these songs are reused.

**Avatar**. Hypothesis *H5* (*customizing the profile avatar is positively associated with receiving overdubs*) could not be tested for recurring collaborations because all the Songtree authors in this dataset use a custom avatar. Regarding occasional collaborations, as shown in Table 7, the hypothesis is confirmed since we found that changing the default avatar (i.e., `has_avatar=True`) is significantly associated with higher odds of song remixing (+17%, p<.001) as well as a higher expected count of overdubs received (20%, p<.001).

**Song specs**. Regarding $\overline{H}6$ (*song metadata tags are positively associated with receiving overdubs*), we find mixed support. Applying tags is not significantly associated with the reuse of songs created in recurring collaborations (see Table 6) whereas in the case of occasional collaborations the `has_tag` predictor is positively associated only with the odds of songs being overdubbed at all (+126%, see Table 7). Since tags are mostly used to highlight needed instruments, we speculate that is not necessary in the case of recurring collaborations, where authors are arguably more familiar with the skills and needs of recurring collaborators. We find that most songs in the dataset contain tags related to the voice track (wanted or already present) but only a few contain tags for other instruments. Therefore, we also speculate that, overall, tagging songs might be more strongly associated with remixing if instrument-related tags were used more often by the community.

**Frequent interaction.** For $\overline{H}7$ (*overdub invitations are positively associated with receiving overdubs*), we find mixed results. For recurring collaborations (see Table 6), we observe that a log-unit increase in `#invitations` sent by authors is associated with a 22% increase in the odds of songs being overdubbed at all. Albeit it was reasonable to expect a significant and positive association of this predictor with song remixing overall, this result is in contrast with the findings from the regression analysis on occasional collaborations, where we observe a lack of significance for the predictor (see Table 7). This contrasting finding suggests that overdub invitations sent to and from frequent collaborators may not be overlooked as in the case of occasional collaborations.

We find mixed results also for $\overline{H}8$ (*the amount of communication exchanged between two authors is positively associated with exchanging overdubs*). Concerning recurring collaborations, we find both `#sent_messages` and `#received_messages` to be positive and significant. Specifically, a log-unit increase in each predictor is associated respectively with a 20% and 34% increase in the odds of songs receiving at least one overdub; at the same time, a log-unit increase in the number of messages sent and received by its author is associated with a 3% increase in the expected count of `#overdubs` received by a song. These results are in contrast with the lack of significance of the `#invitations` predictor in the regression analysis of occasional collaborations – another piece of evidence that the interaction between frequent collaborators is useful to foster song remixing.

> **Remix factors in Songtree collaborations.** Songs receiving positive feedback as well as fresh and less mature songs are generally more likely to be remixed. Unlike songs generated through occasional collaborations, exchanging messages and invitations to collaborate are positively associated with remixing songs generated through recurring collaborations; instead, the opposite is true for reputation since ranking is positively associated only with reusing songs in occasional collaborations. Finally, authors unlocking the badges for receiving overdubs start a virtuous circle that fosters the reuse of their songs.

## 7 DISCUSSION

In this section, we discuss our results as compared to prior research and show their practical implications for Songtree users who want to act and improve their social ranking, as well as online-community designers who aim at improving the collaborative aspects of their music platforms.

### 7.1 Recurring Collaborations

Our analysis revealed that recurring collaborations in Songtree are real. After applying our frequent pattern mining algorithm, we were able to uncover 43 instances of recurring collaborations, of which 5 are in-presence (see the taxonomy in Sect. 6.1). The online-only recurring collaborations are 36, of which 2 are classified as established. While the expert consultation session increased the confidence in the validity of these findings, it was also useful to identify a couple of examples of in-presence bands that we had failed to uncover because of the lack of any references to their existence in the members' username and bio. Therefore, in a way, we can claim that, albeit precise, our approach has failed to recall all the existing instances of recurring collaborations due to the inherent limitation of a solution based on keyword spotting.

In addition, we found out that: (i) all the online-only recurring collaborations except one are formed by two members; (ii) the members of recurring collaborations have a very similar ranking in the community; (iii) the music genre preferences, as well as the set of instruments played, vary when comparing their solo activity to the activity as a member of a recurring collaboration. Regarding the first point, we discussed with the Songtree founder whether this is a side effect of the dyadic nature of overdubbing. However, he dismissed this speculation because he looks at collaboration in Songtree rather at the tree level and, therefore, was more inclined to interpret the finding as a side effect of the lack of more advanced collaborative features such as project and file management, and others typically available in software development platforms. Another possible explanation is an intrinsic limitation of our frequent pattern mining approach based on frequencies. Despite the expert validation session, we cannot ensure that our results are complete—i.e., there might be more recurring collaborations possibly consisting of more than two or three members. Frequency-based models for frequent pattern mining are very popular because of their simplicity. However, other approaches exist such as those based on graphs [55], which might be able to identify other instances of recurring collaborations. We reserve to further investigate the phenomenon of recurring collaborations with graph-based approaches in future work. Still, we point out that this finding about the number of members involved in recurring collaborations is in line with previous work that also found pairwise collaborations to be extremely more prevalent in the FAWM music community [19, 52]. Also, similar findings have been reported in [20] and [57] concerning the size of putative sub-teams of developers in the Apache OSS ecosystem. As for the second point, we speculate that the preference for working with fellow authors sharing the same level of expertise is because recurring collaborations are not a means for experienced musicians to teach newbies but rather a way to help each other and keep growing together. Concerning the third point, we speculate that the different member habits in occasional and recurring collaborations as compared to solo activities are because occasional collaborations are used to explore and experiment with types of music and instruments different from those usually played, whereas recurring ones are fostered by the lack of some skills in the other party.

Finally, with our novel frequent pattern mining algorithm, we have established the *minsup* threshold associated with these recurring collaborations (i.e., to filter out spurious, non-frequent instances), thus providing the first reference value for future research on frequent pattern mining applied to the domain of online music co-creation.

*7.2 Remix Factors*

Table 8 lists the eight hypotheses tested in the study and whether we found support for them when analyzing both occasional collaborations (see column (*a*)) and recurring collaborations (see column (*b*)). The other columns (*c*) and (*d*) show whether we found support for some of the hypotheses in our previous work where we used, respectively, a smaller dataset and a simpler regression model [7] and compared the antecedents of reuse in Songtree to Splice and ccMixter, two other platforms for online music co-creations [8]. The last column (*e*) presents related results from prior work investigating other types of creative communities. Next, we discuss and compare these findings.

**Popular songs**. The overall results of the first hypothesis *H1* show that, except for `#comments`, popular songs that receive positive feedback and appreciation (i.e., `#likes`, `#plays`, `#bookmarks`, and `#reposts`) have significantly higher odds of being overdubbed and receiving significantly more overdubs than the others (see columns (*a-c*)). Consistent results have been observed in our previous work [8] studying Splice and ccMixter (see column (*d*)). These findings hold also in other types of communities and platforms, such as Stack Overflow, for which Calefato et al. [9] found that the number of upvotes received by an answer is strongly associated with its likelihood of being accepted as a solution (see column (*e*)). At the same time, these findings complement the results from previous studies performed on other creative arts communities, such as Newgrounds [33,34] and FAWM [6,46], which did not evaluate the popularity of creative artifacts as a predictor of future successful collaborations. The comparison with previous work on OSS communities is more difficult because there is no such thing as 'popular' pull requests or patches: Tsay et al. [51] studied pull request acceptance in GitHub, discussing the association with popularity at the project level and using the number of stars and collaborators as a proxy.

Table 8. The eight hypotheses tested in the study: column (*a*) shows the result of the regression on occasional collaborations whereas column (*b*) on recurring collaborations; column (*c*) lists the results from our prior work using a different regression model and a smaller dataset; column (*d*) lists the results from our prior work on other online communities for music co-creation; column (*e*) lists results from prior work investigating other types of creative communities.

| Hypothesis | Supported? | | | | |
|---|---|---|---|---|---|
| | This study | | Prior work | | |
| | (*a*) Occasional collaborations | (*b*) Recurring collaborations | (*c*) Calefato et al. [7] | (*d*) Calefato et al. [8] | (*e*) Others |
| *H1* The number of reactions generated by songs is positively associated with receiving overdubs | ~Partially | ~Partially | ~Songtree | ~Splice ~ccMixter | ✓Stack Overflow [9] |
| *H2* Time is negatively associated with receiving overdubs | ~Partially | ~Partially | ✓Songtree | ✗Splice ✗ccMixter | - |
| *H3* The degree of derivativity of songs is negatively associated with receiving overdubs | ✓Yes | ✓Yes | ✓Songtree | ✓Splice ✓ccMixter | ✗Scratch [28] ✓ccMixter [11] |
| *H4* The ranking of authors in the community is positively associated with receiving overdubs | ~Partially | ~Partially | ~Songtree | ✗Splice ~ccMixter | ✓Scratch [28] ✓Wikipedia [1],[26] |
| *H5* Customizing the author profile avatar is positively associated with receiving overdubs | ✓Yes | N/A | ✓Songtree | N/A Splice ✓ccMixter | ✓GitHub OSS communities [24] ✓Newgrounds [33,34] |
| *H6* Applying technical metadata tags to songs is positively associated with receiving overdubs | ~Partially | ✗No | - | - | ✓Wikipedia [45] ✓Newgrounds [35] |
| *H7* Sending overdub invitations is positively associated with receiving overdubs | ✗No | ~Partially | - | - | - |
| *H8* The amount of communication exchanged between two authors is positively associated with exchanging overdubs | ✗No | ✓Yes | - | - | ✓Newgrounds [34] ✓Wikipedia [26] ✓Python OSS community [16] ✓GitHub [31] |

**Recent and mature songs**. Regarding *H2*, we found evidence that the longer since the upload of a song, the fewer its chances of being overdubbed. Also, we observed similar results when studying the ccMixter and Splice music communities [8], thus reinforcing the soundness of the finding that song novelty is a strong antecedent of remix. However, the more sophisticated count data model developed for this study helped clarify that the association is only with the odds of songs being overdubbed at all. Instead, the expected count of overdubs received is positively associated with time, that is, remixed songs seem to benefit from a cumulative advantage whereby "*works exhibiting a high degree of reuse become more attractive for further reuse*" ([13], p. 168).

Regarding *H3*, we hypothesized that the more distant an overdub is from the root of its song tree (i.e., the more derivative it is), the closer it gets to being considered 'finished.' Hence, more polished songs might be perceived as harder to remix and less stimulating. The results of the regression analyses confirmed that more mature, 'complete' songs created towards the end of a long collaboration process are less likely to be overdubbed (see columns (*a,b*) in Table 8). Besides, this finding is not only consistent across platforms (i.e., in ccMixter and Splice, see column (*d*)), but also confirms the observation reported by Cheliotis et al. [11] about the inverse relationship between the degree of generativity and derivativity of music artifacts (see column (*e*)). According to Zittrain [59], generativity in online technologies is indeed linked to the notions of incompleteness and early-stage release, which lead to eliciting more contributions due to the users' perception of increased creativity and simplified participation. This result, however, is in contrast with the finding of Hill and Monroy-Hernandez [28], who found that reused interactive media in the Scratch community are more generative than *de novo* content. These contrasting results about generativity and derivativity are arguably explained by the different types of artifacts, suggesting that reused songs may lose generativity faster than animations—in other words, there may be more ways to expand the story behind an animation than instruments to add to a song.

**Reputation and author profiles**. Prior work (see column (*e*)) has reported on the positive association between social ranking and artifact reuse in arts communities [13,30,38] as well as OSS [34]. Sinnreich [49] found that remixing is driven by the will to create connections with salient creators. Consistently, Hill and Monroy-Hernández [28] and Cheliotis et al. [11] found that authors' prominence and their social embeddedness in the Scratch community are associated with an increase in the likelihood of remixing. Likewise, Halfaker et al. [26] found that user reputation is a strong factor predicting whether additions to Wikipedia pages will stick. In addition, Jiang et al. [31] studied forked repositories in GitHub and found that developers fork more often those owned by 'attractive' (i.e., popular) ones. Instead, the results of our study (see columns (*a,b*)) only provide partial support for the hypothesis *H4* that the reputation of authors in the community—operationalized as the number of followers, ranking, and earned badges—is positively associated with higher odds of their songs being overdubbed as well as a higher expected count of overdubs received. We notice that ranking is not significant in the case of recurring collaboration, arguably because their members have small deltas in community status (see Table 3)—a confirmation of the existing evidence by Settles and Dow [46] who found that a similar status in the FAWM music community was a key factor in pairing members, ensuring the perception of balanced efforts, and completing collaborations. The results about `ranking` were mixed also in our previous work [8] (see column (*d*)) where we found partial support in the ccMixter community and no support in Splice. Regarding badges, the findings are consistent for both occasional and recurring collaboration, yet mixed: our interpretation is that the badges which reflect being a productive community member (i.e., `new_songs_badges` and `overdub_badges`) are not significant, unlike those (i.e., `overdub_received_badges`) that are earned through appreciation (remixes) received from other community members.

Furthermore, we found support for hypothesis *H5*, according to which songs of authors easily recognized by their profile picture are overdubbed more. Our finding confirms our intuition that Songtree members perceive the effort put into curating their personal space as a proxy of the attention put into creating their music. According to Postmes et al. [42], customizing personal information such as the avatar is a form of self-disclosure and self-presentation that shifts the attention from the value of the whole community to the individuals and their activity. This result holds across platforms too (see column (*d*) in Table 8) and is in line consistent with the evidence from prior work (see column (*e*)): Gousios et al. [24] found that the identity of pull request submitters in OSS communities is a very significant predictor for assessing the quality of code contributions. Besides, Luther et al. [33,34] found that being able to browse members' history of contributions is associated with an increase in the chance of successfully completing collaborations. Still, we argue that the role of identity and reputation as predictors of artifact reuse may vary in OSS and arts communities. Code changes in OSS communities need to be reviewed and approved before being integrated with the existing codebase. In Songtree and, more in general, in arts communities there is no counterpart to code review – i.e., 'bad' remixes get in as well as 'good' ones. As such, the significance of identity and reputation as proxies of contribution quality in arts communities may be weaker. We reserve to further investigate this comparison in future work. Still, given the overall results for $\overline{H}7$ and $\overline{H}8$, our study provides further evidence that the social ranking of users in the arts community is positively associated with fostering artifact reuse.

**Song specs**. We found mixed results regarding the hypothesis $\overline{H}6$ that tagging songs with technical specifications is associated with an increase in the chances of being overdubbed. Specifically, unlike songs generated through recurring collaborations (see column (*b*)), for those generated through occasional collaborations (column (*a*)) applying technical tags is positively associated with the odds of being overdubbed at all. As regards prior work (see column (*d*)), our findings are somewhat in line with those reported by Luther et al. [35] (see column (*e*)) who found that Newgrounds animations advertising technical specifications are associated with higher chances of reuse. Also, prior work on Wikipedia (e.g., [1,30,45,54]) has consistently found evidence that article metadata in Wikipedia are reliable proxies for article quality and, as such, they can be used as antecedents of page edits that will stick. Our inspection revealed that most of the tags concern vocals as compared to others such as the instruments played or still missing, music key, and tempo. We speculate that this difference may be because recurring collaborators do not need to look at the technical tags to know the others' technical preferences and skills possessed or lacked.

**Frequent interactions**. Regarding $\overline{H}7$, we found that invitations sent by recurring collaborators (see column (*b*)) are positively associated with the increased odds of songs being overdubbed at all. Instead, the hypothesis was not supported for songs generated through occasional collaborations (column (*a*)). Consistent findings were found regarding $\overline{H}8$ as we identified a positive association between the exchange of messages with recurring collaborators and both the odds of overdubbing each other's songs and the expected count of overdubs received by their songs. According to the common bond theory [37,44], in fact, frequent communication is key to creating and maintaining strong online relationships. Overall, these results are consistent with prior research (see column (*e*)). Ducheneaut [16] found that when authors of external contributions to OSS projects have previously interacted with project team members, they have higher chances to have their source code extensions integrated. Also, Luther et al. [34] found that frequent communication is a common success factor in both OSS and arts communities. Nonetheless, we speculate that the lack of significance for occasional collaborators' interaction is because receiving overdub invitations may feel like 'cold calls' until the parties get acquainted. We reserve to further investigate this aspect in future work as in this study we have not been able to assess the link, if any, between the *minsup* threshold before collaborations are considered frequent and the amount of communication needed before overdub invitations are not perceived as 'cold' anymore.

**Interpretation and synthesis of results**. Songtree is an example of a collaborative social platform. Collaborating with others means being on each other's radar and, consequently, using digital signals to make sense of the skills and qualities possessed by fellow musicians. To that end, we argue that the *signaling theory* [12] is a useful framework to understand what pieces of information are more reliable and consequently explain why some remix factors drive more than others the song reuse behavior in Songtree.

Regardless of the application domain, the signaling theory posits that signals that are costly to fake for the sender are also the most reliable. In the digital realm, Donath [14] distinguishes two types of signals: (i) *assessments signals*, which relate to the qualities possessed by the sender that can be directly assessed by observing them (e.g., the record of positive transactions in an online marketplace signals sellers' reliability); (ii) *conventional signals*, which instead are not correlated directly to a sender's trait and, therefore, are potentially deceiving (e.g., a retouched profile picture on a dating site). According to the signaling theory framework, Songtree users use positive reactions received by songs, user ranking, and overdubs received badges as assessment signals (i.e., costly to fake) to make sense of authors' music skills; likewise, they look at the effort spent into applying tag specifications and customizing their avatars as a signal of commitment. Instead, the theory suggests that the new songs badges and overdub badges are not significant remix factors because they send conventional signals that are potentially deceiving as authors may intentionally inflate them by uploading lots of poor-quality new songs and overdubs.

The signaling theory framework is also useful to provide a synthesis of the similar and contrasting findings from prior work. Specifically, we speculate that these differences may be (also) related to the different types of artifacts created in these platforms (e.g., song tracks in Songtree and ccMixter, animations in Newgrounds, text in Wikipedia) because of the distinct signals that they can convey. Consider H1 for example: unlike overdubs, source code additions in GitHub and page edits in Wikipedia do not receive appreciation feedback *per se*—unlike the entire repositories and pages (via the number of stars and visits)—and, hence, they are unable to carry easy-to-access proxy signals for the assessment of the author's skills. The assessment is still possible but, for example, only upon a time-consuming code review. Therefore, we argue that some contrasting findings are not only to be expected but also intrinsic to the different platforms and artifacts under study.

*7.3 Practical Implications*

Building on our findings, here we first propose some practical recommendations addressed to music platform designers to improve the collaborative aspects of such platforms; then, we propose recommendations addressed to Songtree users who want to improve their status in the community.

**Design Recommendations for Music Co-creation Platform Designers**. During our interactions with the Songtree developers, they reported that users sometimes complain about the poor quality of some of the overdubs added to a song tree they started. Inspired by code reviews performed in software development environments, the designers of collaborative music platforms should consider implementing optional, pull request-like review mechanisms for accepting remixes and preventing low-quality extensions, instead of relying on features such as closing and hiding songs altogether as in Songtree.

Also inspired by collaborative development environments, we call for adding support to bands in online music co-creation platforms. Interestingly, the request for adding support for online-only bands had already emerged in our previous work [8] where a few study participants called for implementing features that would allow Songtree users to "*set a virtual band [and] group their songs together,*" assign roles explicitly (e.g., producer), and possibly rely on a more sophisticated chat system. Building on the feedback gathered during the interview with the Songtree founder, to ease the formation of online-only bands platform designers should consider implementing features that match the profiles of musicians seeking to start a new band and, to facilitate bonding [41], recommend prospective members to existing bands who are missing a specific skill or have a similar music taste. Also, to further attract existing in-presence bands to their platform, designers should consider adding more sophisticated features to explicitly support collaboration, such as project and file management, multi-track mixing. Yet, they should also be aware of the tension between adding such sophisticated features to support collaboration among more expert musicians and the simplicity of overdubbing (i.e., two-track mixing), which facilitates more amateur users.

Another practical recommendation is to further leverage the positive association of author prominence with an increase in the likelihood and count of remixes. Specifically, online music platform designers should leverage the signaling theory framework to increase the visibility of the assessment signal related to reputation, for which the regression analyses revealed positive associations in Songtree, i.e., the number of followers, ranking, and the overdubs received badges, and at the same time demote those for which we found a negative association, i.e., the badges earned by uploading new and remixed songs.

Regarding tags, we noticed that voice-related metadata tags are applied consistently to songs whereas other technical tags are seldom used. As such, to increase their usefulness, online music platform designers should consider adopting some sort of gamification mechanism to encourage authors to consistently annotate songs with metadata tags.

Finally, we provide a couple of recommendations that are specific to Songtree. Our regression analyses revealed partial support for the hypotheses related to sending invitations to overdub. After looking into the implementation of the feature in Songtree, we argue that overdub invitations can go unnoticed because they are not separated from other types of notification. Accordingly, Songtree designers should consider overhauling the notification system to highlight overdub requests and other important notifications that should not be missed. Also, the Songtree maintainers should consider implementing a recommender system that introduces some randomness or other factors to favor the discovery of 'old but gold' songs.

**Actionable Recommendations**. Some of the findings from our study are particularly relevant because they are actionable. In particular, users seeking to increase the number of remixes received, and consequently climb a community's social rank, should consider either starting new songs or reusing less mature ones so that others will still have enough 'room' to build upon their work.

A second practical recommendation for those seeking to increase their reputation is to invest time in keeping in touch with fellow authors with whom they interact regularly (i.e., their recurring collaborators) rather than reaching out to many occasional ones.

Finally, where applicable, authors should consider investing their time in applying tags to annotate the technical specifications of songs to facilitate their reuse.

*7.4 Limitations*

One possible limitation of this work concerns the external validity (i.e., generalizability) of our findings as one can argue that Songtree is not representative of all music co-creation platforms and, therefore, that our results might not transfer to other communities. However, as highlighted in Table 8, we have already verified that many findings related to reusing song tracks in Songtree successfully transfer to other songwriting communities (see column (*d*)) as well as other types of creative communities (see column (*e*)).

We also identified a few limitations that affect construct validity, which concerns the degree of accuracy to which the variables (i.e., features) measure the constructs of interests. In this study, we have collected cross-sectional data, which does not allow us to clear the causality nexus. However, the availability of the entire dataset dump of Songtree gave us access to the entire history of events and, thus, made it possible to mitigate reverse causality issues by extracting the predictors just before the event of interest is observed. Therefore, albeit in the formulation of the hypotheses and discussion of the results we have hypothesized and identified positive/negative associations, we are also certain about the underlying direction of causality between the dependent variable in the count data models and the occurrence of any of the predictors. Nonetheless, in future work, we will collect further snapshots of the Songtree database and perform longitudinal analyses that will allow us to make stronger inferences about causality.

In our regression analyses, we investigated remix factors related to the upload time, feedback received by songs, and author reputation. We acknowledge that presence in Songtree of sections such as *Latest*, *Popular*, and *Top artists*, featuring recent as well as prominent songs and artists, represents a confounding factor that may raise rival explanations to the findings of our study on the antecedents of song reuse. However, in prior work, Stanko [50] found that promoting artists and songs on the front pages of the Thingiverse community website was not associated with the likelihood of remixing.

Regarding the `has_avatar` construct, which we defined as a proxy for author profiles that are easier to recognize, we acknowledge that users may unintentionally select custom profile images that are nevertheless hard to recognize. We also acknowledge that using only a dichotomous variable to model an entire author's identity is a simplistic operationalization that shall be addressed in future replications.

Concerning the ranking construct, Cheliotis et al. [11] operationalized the prominence of authors in an online creative community using a bow-tie analysis. Instead, we relied on the same metric of commitment defined by Songtree designers. In a future extension, we aim to leverage clique analysis for a finer-grain evaluation of authors' social ranking as an overdub factor.

Finally, regarding the mining of recurring collaborations, we acknowledge that conducting interviews with the musicians involved in collaborations would have been a better approach to validate the findings. However, this option was not allowed as per the signed NDA agreement. Nonetheless, the expert validation session conducted with the Songtree founder was sufficient to find out that the approach based on string matching for identifying established, online-only collaborations and in-presence bands is not 100% complete, albeit precise (i.e., no false positives). Considering that Songtree is currently lacking specific features to support bands, it appears reasonable that stronger collaborations may move off the platform and use other channels (e.g., SoundCloud). Accordingly, we acknowledge the lack of cross-platform data in the present study and aim to fix this limitation in future work. Also, we found that all recurring collaborations involve two or three user accounts; however, we cannot exclude cases where any of these accounts is used by multiple people at the same time.

## 8 CONCLUSIONS

In this paper, we conducted a two-stage study in the Songtree music community to analyze in-depth the remix factors of occasional and recurring collaborations. In each stage, we created a statistical model to validate the association of several song- and author-related antecedents of reuse with the likelihood of songs being overdubbed (i.e., extended by another musician) as well as the overall count of overdubs received.

Overall, regarding the song-related remix factors, we found that both recent and less mature songs, as well as those that generate many reactions (e.g., likes, plays, bookmarks), are more likely to be derived while also receiving a higher number of overdubs. We also found mixed evidence about the association of applying technical specifications (tags) with the likelihood of receiving overdubs and the expected number of times a song is reused. Concerning the author-related factors, we found that popular authors–i.e., highly ranked and with many followers and remixes– have higher odds of seeing their songs further remixed.

As regards recurring collaborations, we developed an algorithm based on frequent pattern mining and uncovered 38 online-only collaborations, all composed of two or three members with a similar ranking in the community. As compared to occasional collaborations, direct messaging and invitations to remix are significant antecedents of reuse only for frequent collaborators whereas the opposite is true for author ranking.

We compared our results with prior work on both OSS and online artistic communities to highlight common factors that generalize across music co-creation platforms and beyond the music domain. We also derived actionable recommendations to Songtree users seeking visibility and practical recommendations to inform the designers of creative arts communities about aligning their policies with the assessment signals that members are already using to infer authors' skills from the music artifacts.

In future work, we intend to leverage social network analysis to analyze the collaboration and communication networks in Songtree. We are also looking at graph-based frequent pattern mining approaches to uncover potentially more and larger recurring collaborations in Songtree. Finally, as we acquire further snapshots of the database, we intend to conduct a longitudinal study to uncover new antecedents of song reuse, such as the retention and loyalty of community members over time.

## ACKNOWLEDGMENTS

We thank Songtree and, in particular, Flavio Antonioli, for providing the data and supporting us during our study. We also thank Claudia Capozza for the feedback on the regression analysis.

An in-depth Analysis of Occasional and Recurring Collaborations    32[20] Mohammad Gharehyazie and Vladimir Filkov. 2017. Tracing distributed collaborative development in apache software foundation projects. *Empir. Softw. Eng.* 22, 4 (August 2017), 1795–1830. DOI:https://doi.org/10.1007/s10664-016-9463-3

[21] Lucy L. Gilson, M. Travis Maynard, Nicole C. Jones Young, Matti Vartiainen, and Marko Hakonen. 2015. Virtual Teams Research. *J. Manage.* 41, 5 (July 2015), 1313–1337. DOI:https://doi.org/10.1177/0149206314559946

[22] Georgios Gousios, Martin Pinzger, and Arie van Deursen. 2014. An exploratory study of the pull-based software development model. In *Proceedings of the 36th International Conference on Software Engineering - ICSE 2014*.

[23] Georgios Gousios, Margaret-Anne Storey, and Alberto Bacchelli. 2016. Work practices and challenges in pull-based development: The contributor's perspective. In *Proceedings of the 38th International Conference on Software Engineering - ICSE '16*.

[24] Georgios Gousios, Andy Zaidman, Margaret-Anne Storey, and Arie van Deursen. 2015. Work Practices and Challenges in Pull-Based Development: The Integrator's Perspective. In *2015 IEEE/ACM 37th IEEE International Conference on Software Engineering*.

[25] William H. Greene. 2002. *Econometric Analysis* (5th ed.). Prentice Hall. DOI:https://doi.org/10.1007/978-3-540-78389-3

[26] Aaron Halfaker, Aniket Kittur, Robert Kraut, and John Riedl. 2009. A jury of your peers: quality, experience and ownership in Wikipedia. In *Proceedings of the 5th International Symposium on Wikis and Open Collaboration - WikiSym '09*, ACM Press, New York, New York, USA, 1. DOI:https://doi.org/10.1145/1641309.1641332

[27] Lieve Hamers, Yves Hemeryck, Guido Herweyers, Marc Janssen, Hans Keters, Ronald Rousseau, and André Vanhoutte. 1989. Similarity measures in scientometric research: The Jaccard index versus Salton's cosine formula. *Inf. Process. Manag.* 25, 3 (January 1989), 315–318. DOI:https://doi.org/10.1016/0306-4573(89)90048-4

[28] Benjamin Mako Hill and Andrés Monroy-Hernández. 2013. The Remixing Dilemma. *Am. Behav. Sci.* 57, 5 (May 2013), 643–663. DOI:https://doi.org/10.1177/0002764212469359

[29] Mei-Chen Hu, Martina Pavlicova, and Edward V Nunes. 2011. Zero-inflated and hurdle models of count data with extra zeros: examples from an HIV-risk reduction intervention trial. *Am. J. Drug Alcohol Abuse* 37, 5 (September 2011), 367–75. DOI:https://doi.org/10.3109/00952990.2011.597280

[30] Sara Javanmardi, David W. McDonald, and Cristina V. Lopes. 2011. Vandalism detection in Wikipedia. In *Proceedings of the 7th International Symposium on Wikis and Open Collaboration - WikiSym '11*, ACM Press, New York, New York, USA, 82. DOI:https://doi.org/10.1145/2038558.2038573

[31] Jing Jiang, David Lo, Jiahuan He, Xin Xia, Pavneet Singh Kochhar, and Li Zhang. 2017. Why and how developers fork what from whom in GitHub. *Empir. Softw. Eng.* 22, 1 (February 2017), 547–578. DOI:https://doi.org/10.1007/s10664-016-9436-6

[32] J. Scott Long and Jeremy Freese. 2014. *Regression Models for Categorical Dependent Variables Using Stata* (3rd ed.). Stata Press, College Station, Texas. Retrieved June 19, 2021 from https://www.stata.com/bookstore/regression-models-categorical-dependent-variables/

[33] Kurt Luther and Amy Bruckman. 2008. Leadership in Online Creative Collaboration. *Proc. 2008 ACM Conf. Comput. Support. Coop. Work* (2008), 343–352. DOI:https://doi.org/10.1145/1460563.1460619

[34] Kurt Luther, Kelly Caine, Kevin Ziegler, and Amy Bruckman. 2010. Why It Works ( When It Works ): Success Factors in Online Creative Collaboration. *Assoc. Comput. Mach.* 10, (2010), 1–10. DOI:https://doi.org/10.1145/1880071.1880073

[35] Kurt Luther, Kevin Ziegler, Kelly E. Caine, and Amy Bruckman. 2009. Predicting successful completion of online collaborative animation projects. *Proceeding seventh ACM Conf. Creat. Cogn. - C&C '09* (2009), 391. DOI:https://doi.org/10.1145/1640233.1640316

[36] Sanna Malinen. 2015. Understanding user participation in online communities: A systematic literature review of empirical studies. *Comput. Human Behav.* 46, (May 2015), 228–238. DOI:https://doi.org/10.1016/j.chb.2015.01.004

[37] Katelyn Y.A. McKenna, Amie S. Green, and Marci E.J. Gleason. 2002. Relationship formation on the internet: What's the big attraction? *J. Soc. Issues* 58, 1 (January 2002), 9–31. DOI:https://doi.org/10.1111/1540-4560.00246

[38] Audris Mockus. 2009. Succession: Measuring transfer of code and developer productivity. In *Proceedings - International Conference on Software Engineering*, 67–77. DOI:https://doi.org/10.1109/ICSE.2009.5070509

[39] Samuel Müller, J. L. Scealy, and A. H. Welsh. 2013. Model Selection in Linear Mixed Models. *Stat. Sci.* 28, 2 (May 2013), 135–167. DOI:https://doi.org/10.1214/12-STS410

[40] Eduardo Navas. 2012. *Remix Theory: The Aesthetics of Sampling*.

[41] So Yeon Park and Blair Kaneshiro. 2021. Social Music Curation That Works: Insights from Successful Collaborative Playlists. *Proc. ACM Human-Computer Interact.* 5, CSCW1 (April 2021), 1–27. DOI:https://doi.org/10.1145/3449191

[42] Tom Postmes, Russell Spears, and Martin Lea. 2002. Intergroup differentiation in computer-mediated communication: Effects of depersonalization. *Gr. Dyn.* 6, 1 (2002), 3–16. DOI:https://doi.org/10.1037/1089-2699.6.1.3

[43] J. Preece. 2000. *Online communities: Designing usability, supporting sociability*. Wiley, New York, NY, USA.

[44] Yuqing Ren, Robert Kraut, and Sara Kiesler. 2007. Applying common identity and bond theory to design of online communities. *Organ. Stud.* 28, 3 (March 2007), 377–408. DOI:https://doi.org/10.1177/0170840607076007

[45] Jeffrey Segall and Rachel Greenstadt. 2013. The illiterate editor. In *Proceedings of the 9th International*

**Appendix A – Descriptive statistics of measures from the occasional collaboration dataset**

| | Measure | Total | | Min | Max | Mean | St. Dev. | Variance |
|---|---|---|---|---|---|---|---|---|
| | | Yes | No | | | | | |
| Song level | #likes | 1,632,978 | | 0 | 1,459 | 7.46 | 63.4 | 4019.12 |
| | #bookmarks | 306,424 | | 0 | 184 | 1.4 | 12.39 | 153.47 |
| | #plays | 50,494,162 | | 0 | 55461 | 230 | 2,355.25 | 5,547,200 |
| | #reposts | 204,026 | | 0 | 173 | 0.93 | 7.54 | 56.85 |
| | #comments | 292,747 | | 0 | 117 | 1.34 | 6.35 | 40.28 |
| | upload_time_interval | - | | 2 | 2,391,769 | 738,714 | 29,492,939 | 8.698334e+14 |
| | song_depth | - | | 0 | 19 | 0.36 | 0.94 | 0.89 |
| | has_tags | 218,368 | 534 | - | - | - | - | - |
| | #invitations | 2,184,780 | | 0 | 2,396 | 9.98 | 82.14 | 6,747.33 |
| Author level | msg_exchange_rate | - | | 0 | 356,257 | 65 | 2,643.77 | 6,989,495 |
| | #followers | 12,278,941 | | 0 | 668 | 56.1 | 128.78 | 16,584.14 |
| | ranking | - | | 0 | 22,680 | 47.55 | 293.15 | 85,939.7 |
| | new_songs_badge | Rookie 30,324 Songwriter 52,249 Composer 22,547 | 113,782 | - | - | - | - | - |
| | overdubs_badge | Performer 13,294 Top performer 19,497 Virtuoso 28,781 | 157,330 | - | - | - | - | - |
| | overdubs_received_badge | Songsmith 12,980 Band leader 22,174 Maestro 32,470 | 151,278 | - | - | - | - | - |
| | has_avatar | 197,324 | 21,578 | - | - | - | - | - |

**Appendix B – Regression model selection strategy**

In modeling count data, we follow the approach suggested by Cameron and Trivedi [10] and Green [25]. First, we considered the Poisson regression model. The Poisson distribution assumes equidispersion, that is the equality of mean and variance of the count-dependent variable. However, count data frequently depart from the Poisson distribution due to overdispersion, that is, a larger frequency of extreme observations resulting in spread (variance) greater than the mean in the observed distribution. As such, if the dependent variable is overdispersed (i.e., its variance exceeds its mean), the Poisson regression model may lead to inconsistent estimates. In such cases, count data can be modeled using the negative binomial distribution, a generalization of Poisson distribution, which adds a parameter to accommodate for the overdispersion. The negative binomial distribution converges to the Poisson distribution if the overdispersion parameter tends to zero. As shown by the descriptive statistics reported in Appendix A, the number of overdubs received by songs in the experimental dataset has a mean equal to 1 (SD=32.1) and variance of 1,033. This is an indication of overdispersion. Thus the negative binomial model is preferred to the Poisson model. A formal Likelihood Ratio Test (LRT) of overdispersion is performed to ascertain that the negative binomial model provides a better fit to the data than the Poisson model, that is, the null hypothesis of equidispersion (Poisson model) is tested against the alternative of overdispersion (negative binomial model).

Second, the distribution of counts often exhibits several observed zeros larger than what is assumed by the Poisson distribution. Zero-inflated and hurdle models [29] have been developed to cope with the high occurrence of zeros in the outcome data, whether overdispersed (negative binomial) or not (Poisson distribution). Both zero-inflated and hurdle are two-part models. The first part is a binomial probability (i.e., logistic regression) model that determines whether a zero or non-zero outcome occurs. This logistic regression allows us to study why some songs are not overdubbed while others are. The second part is a zero-truncated count data distribution (either Poisson or negative binomial, depending on the LRT above), which models the positive outcomes. This regression allowed us to understand why some songs receive a higher number of overdubs than others.

The standard analysis of model fit for these methods uses both the Vuong test of non-nested model fit [52] and Akaike Information Criterion (AIC) to determine which model fits best. Accordingly, we first performed two Vuong tests to compare both the hurdle model and the zero-inflated negative binomial model against the standard negative binomial model; then, we used the AIC to select the best fitting model between the resulting two.

## Appendix C – Itemset generation algorithm

Here, we introduce in detail the frequent pattern mining algorithm adapted to the music domain and Songtree's specific context. First, we consider every song tree to be a transaction and every user in the community as an item since we want to uncover frequent itemsets of users who collaborate often as members of recurring collaborations. Then, for any given song tree in the dataset, we visit every path from the root to each node: if two or more authors collaborate via overdubbing, they belong to an itemset contained in the transaction corresponding to the given song tree. Figure 5 below reports an example of the $k$-itemsets mined from a tree of height 2, containing five nodes/authors.

In the procedure, the $1$-itemsets (containing only each of the five authors alone) are not relevant since we are interested in recurring collaborations (i.e., frequent itemsets) that, by definition, contain two or more members. Therefore, for $k \geq 2$ we identify six potential frequent $k$-itemsets: four $2$-itemsets and two $3$-itemsets. In Table 9, we show the itemsets and associations rules for the song tree shown in the figure above, assuming n=100 transactions/trees and $minsup$ = 0.05.

These $k$-itemsets are used to generate the association rules (e.g., $\{Author1,…, AuthorN-1\} \Rightarrow \{AuthorN\}$) by selecting the antecedent and consequent as subsets of each $k$-itemset and retaining those with occurrences >3, i.e., we assume any collaboration happening three times or less to be occasional. The table also reports the association rules generated from the $k$-itemsets drawn from the figure. For the sake of simplicity, in the example, we assume a total of n=100 transactions/trees and a $minsup$ threshold of 0.05. As can be observed in the table, there are three $k$-itemsets ($k \geq 2$) occurring more than 3 times. Since the support for them is larger than $minsup$=0.05, we generate the association rules from each itemset and compute the lift scores. Because their lift values are greater than 1, indicating a strong association between the items, the three rules $\{Author1\} \Rightarrow \{Author2\}$, $\{Author2\} \Rightarrow \{Author4\}$, and $\{Author1, Author2\} \Rightarrow \{Author4\}$ are retained.

Table 9. Itemset association rules for Figure 5 (assuming n=100 transact./trees and *minsup*=0.05)

| *k*-itemset | # occurrences (>=3) | sup (#occur/n>=0.05) |
|---|---|---|
| {Author1} | 20 | 0.2 |
| {Author2} | 25 | 0.25 |
| {Author3} | 50 | 0.5 |
| {Author4} | 40 | 0.4 |
| {Author5} | 10 | 0.1 |
| **{Author1, Author2}** | **17** | **0.17** |
| {Author1, Author3} | 2 | - |
| **{Author2, Author4}** | **20** | **0.2** |
| {Author2, Author5} | 1 | - |
| **{Author1, Author2, Author4}** | **15** | **0.15** |
| {Author1, Author2, Author5} | 1 | - |

$$lift(\{Author1\} \Rightarrow \{Author2\}) = \frac{sup(\{Author1, Author2\})}{sup(\{Author1\}) \cdot sup(\{Author2\})} = \frac{0.17}{0.2 \cdot 0.25} = 3.4$$

$$lift(\{Author2\} \Rightarrow \{Author4\}) = \frac{sup(\{Author2, Author4\})}{sup(\{Author2\}) \cdot sup(\{Author4\})} = \frac{0.2}{0.2 \cdot 0.4} = 2.5$$

$$lift(\{Author1, Author2\} \Rightarrow \{Author4\}) = \frac{sup(\{Author1, Author2, Author4\})}{sup(\{Author1\}) \cdot sup(\{Author2\}) \cdot sup(\{Author2\})} = \frac{0.15}{0.2 \cdot 0.25 \cdot 0.4} = 7.5$$

Finally, we show the itemset generation algorithm as pseudocode. The function `itemsetsCreator` below generates the list of itemsets by visiting the paths of each tree in the dataset. For each path of length *l*, the function `subSequences` generates the candidate *k*-itemsets by identifying all the sub-sequences of length 1 ... *l* in the form A-B-C where each element is the author of the song/node. These subsequences may contain redundancies because of self-overdubs (e.g., consider the subsequence A-A-B-C, which would generate candidate *k*-itemsets such as {A, A, B} and {A, A, B, C}). Such redundancies are respectively reduced as {A, B} and {A, B, C} in the function `itemsetsCreator` and added as unique elements to the set `uniqueItemsets`.

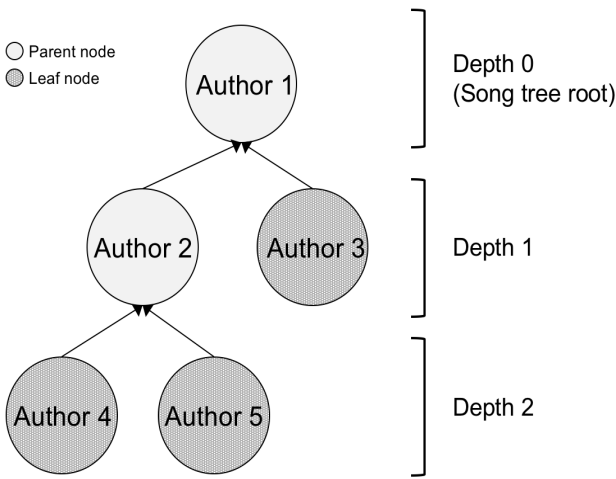

Figure 5. An example of song tree with fine nodes/authors

Finally, we note that the downward closure property is not valid in our domain, due to the nature of the overdubbing relationship between nodes/authors. Consider, for example, a dataset containing only a song tree in the form ABC, from which we extract three itemsets: {A, B}, {B, C}, and {A, B, C}. The itemset {A, B, C} is extracted as frequent even if its subset {A, C} is not. If the downward closure property were valid, all the supersets of {A, C} should be not frequent as well, including {A, B, C}. Hence, in the implementation of the *Apriori* algorithm, the computation of support was computationally expensive as it was not monotonic, and it is not possible to prune the search space. The lack of support for the downward closure property does not invalidate the application of the *Apriori* algorithm to the domain but it increases its computational cost.

**Algorithm** Generate the itemsets from a list of trees

```
 1: function ITEMSETSCREATOR(treeList)
 2:     itemsetList ← []
 3:     for tree ∈ treeList do
 4:         treeItemsets ← []
 5:         uniqueItemsets ← []
 6:         for path ∈ tree do
 7:             treeItemsets.insert(subSequences(path))
 8:         end for
 9:         for itemset ∈ treeItemsets do
10:             itemset ← set(itemset)
11:             if itemset ∉ uniqueItemsets then
12:                 uniqueItemsets.insert(itemset)
13:             end if
14:         end for
15:         itemsetList.insert(uniqueItemsets)
16:     end for
17:     return itemsetList
18: end function
19:
20: function SUBSEQUENCES(path)
21:     allSubsequences ← []
22:     l ← length(path)
23:     for i ← 1 to l do
24:         for j ← 0 to (l − i) do
25:             n ← (j + i − 1)
26:             items ← []
27:             for k ← j to n do
28:                 items.insert(path(k))
29:             end for
30:             allSubsequences.insert(items)
31:         end for
32:     end for
33:     return allSubsequences
34: end function
```

**Appendix D – List of the online-only recurring collaborations identified**

| Id | Min-sup | #Collaborations | #Messages | #Invites | Mean #Likes | Mean Coolness index | Avg #songs | Mean len. | Date (yyyy-mm) First | Last |
|---|---|---|---|---|---|---|---|---|---|---|
| 15 | 0.010 | 22 | 15 | 100 | 13.0 | 19.0 | 129 | 135.0 | 2017-04 | 2019-10 |
| 45 | 0.014 | 30 | 0 | 99 | 3.0 | 1.0 | 142 | 234.0 | 2017-06 | 2018-07 |
| 58 | 0.012 | 26 | 71 | 323 | 6.0 | 12.0 | 166 | 116.0 | 2017-06 | 2018-01 |
| 62 | 0.005 | 11 | 824 | 2 | 1.0 | 0.0 | 44 | 142.0 | 2017-07 | 2018-03 |
| 64 | 0.008 | 17 | 16 | 227 | 6.0 | 1.0 | 128 | 198.0 | 2018-10 | 2019-09 |
| 69 | 0.006 | 13 | 0 | 225 | 22.0 | 9.0 | 170 | 200.0 | 2018-06 | 2019-05 |
| 70 | 0.009 | 19 | 29 | 154 | 11.0 | 7.0 | 263 | 205.0 | 2018-02 | 2018-12 |
| 87 | 0.006 | 13 | 0 | 199 | 0.0 | 0.0 | 73 | 227.0 | 2016-07 | 2017-02 |
| 144 | 0.007 | 16 | 3 | 525 | 2.0 | 8.0 | 424 | 212.0 | 2016-05 | 2018-11 |
| 184 | 0.006 | 13 | 90 | 168 | 11.0 | 5.0 | 178 | 202.0 | 2018-08 | 2019-07 |
| 192 | 0.005 | 11 | 15 | 357 | 9.0 | 28.0 | 189 | 216.0 | 2016-07 | 2017-08 |
| 218 | 0.005 | 11 | 1 | 327 | 3.0 | 8.0 | 420 | 252.0 | 2017-02 | 2018-05 |
| 234 | 0.005 | 11 | 227 | 18 | 7.0 | 5.0 | 226 | 241.0 | 2017-04 | 2018-10 |
| 241 | 0.006 | 13 | 0 | 201 | 9.0 | 13.0 | 40 | 222.0 | 2016-11 | 2019-02 |
| 271 | 0.009 | 20 | 7 | 263 | 5.0 | 28.0 | 225 | 220.0 | 2016-11 | 2017-09 |
| 279 | 0.022 | 47 | 567 | 653 | 8.0 | 38.0 | 269 | 233.0 | 2016-07 | 2018-02 |
| 299 | 0.015 | 32 | 0 | 319 | 14.0 | 23.0 | 114 | 229.0 | 2016-08 | 2019-04 |
| 327 | 0.029 | 62 | 559 | 1 | 2.0 | 34.0 | 142 | 177.0 | 2018-01 | 2018-08 |
| 364 | 0.021 | 46 | 2 | 165 | 6.0 | 22.0 | 254 | 197.0 | 2017-07 | 2018-12 |
| 368 | 0.009 | 20 | 10 | 342 | 6.0 | 29.0 | 221 | 210.0 | 2016-07 | 2017-03 |
| 370 | 0.011 | 23 | 0 | 227 | 7.0 | 5.0 | 211 | 206.0 | 2018-07 | 2018-09 |
| 381 | 0.011 | 24 | 1 | 275 | 4.0 | 34.0 | 223 | 183.0 | 2018-02 | 2018-07 |
| 385 | 0.006 | 13 | 35 | 421 | 10.0 | 29.0 | 196 | 228.0 | 2016-02 | 2018-03 |
| 387 | 0.012 | 27 | 92 | 672 | 5.0 | 10.0 | 439 | 233.0 | 2016-04 | 2019-02 |
| 396 | 0.037 | 80 | 997 | 558 | 8.0 | 13.0 | 211 | 125.0 | 2017-04 | 2019-02 |
| 406 | 0.009 | 20 | 71 | 276 | 8.0 | 1.0 | 89 | 219.0 | 2018-11 | 2019-11 |
| 430 | 0.010 | 22 | 13 | 102 | 3.0 | 1.0 | 101 | 231.0 | 2018-11 | 2019-11 |
| 438 | 0.016 | 34 | 0 | 145 | 2.0 | 22.0 | 200 | 241.0 | 2016-12 | 2017-07 |
| 442 | 0.006 | 13 | 14 | 123 | 11.0 | 28.0 | 187 | 219.0 | 2016-04 | 2017-10 |
| 456 | 0.007 | 15 | 1 | 231 | 5.0 | 9.0 | 117 | 188.0 | 2016-07 | 2018-04 |
| 481 | 0.016 | 34 | 159 | 203 | 8.0 | 5.0 | 188 | 233.0 | 2018-07 | 2019-09 |
| 489 | 0.022 | 47 | 0 | 362 | 0.0 | 0.0 | 114 | 246.0 | 2016-05 | 2017-02 |
| 502 | 0.006 | 12 | 87 | 365 | 8.0 | 13.0 | 157 | 107.0 | 2017-06 | 2018-07 |
| 511 | 0.010 | 22 | 54 | 118 | 5.0 | 4.0 | 177 | 283.0 | 2016-09 | 2019-09 |
| 517 | 0.016 | 34 | 9 | 476 | 11.0 | 20.0 | 233 | 134.0 | 2017-06 | 2019-08 |
| 583 | 0.019 | 41 | 246 | 88 | 4.0 | 3.0 | 282 | 113.0 | 2017-04 | 2019-10 |